\UseRawInputEncoding 
\documentclass[journal]{IEEEtran}
\usepackage{graphicx}
\usepackage{setspace}
\usepackage{xspace,amsthm,amsmath,amsfonts,syntonly}
\usepackage{amssymb}
\usepackage{epsfig} 
\usepackage{epstopdf}
\usepackage{epsf,graphics,graphicx}
\usepackage{cite,bm,color,url,textcomp}
\usepackage{float}
\usepackage{subfig}
\usepackage{algorithmicx,algorithm,algpseudocode}
\usepackage{mathrsfs}
\usepackage{arydshln}
\usepackage{stfloats}

\newcommand{\e}{{\rm E} \, }

\newcommand{\ba}{\begin{array}}
\newcommand{\ea}{\end{array}}
\newcommand{\be}{\begin{displaymath}}
\newcommand{\ee}{\end{displaymath}}
\newcommand{\ben}{\begin{equation}}
\newcommand{\een}{\end{equation}}
\newcommand{\bena}{\begin{eqnarray}}
\newcommand{\eena}{\end{eqnarray}}
\newcommand{\beqa}{\begin{eqnarray*}}
\newcommand{\enqa}{\end{eqnarray*}}

\newcommand{\bc}{\begin{center}}
\newcommand{\ec}{\end{center}}
\newcommand{\bi}{\begin{itemize}}
\newcommand{\ei}{\end{itemize}}
\newcommand{\benu}{\begin{enumerate}}
\newcommand{\eenu}{\end{enumerate}}
\newcommand{\bdes}{\begin{description}}
\newcommand{\edes}{\end{description}}
\newcommand{\bt}{\begin{tabular}}
\newcommand{\et}{\end{tabular}}

\newcommand \thetabf{{\mbox{\boldmath$\theta$\unboldmath}}}

\newcommand \alphabf{\mbox{\boldmath$\alpha$\unboldmath}}
\newcommand \betabf{\mbox{\boldmath$\beta$\unboldmath}}

\newcommand \zetabf{\mbox{\boldmath$\zeta$\unboldmath}}

\newcommand \xibf{\mbox{\boldmath$\xi$\unboldmath}}

\newcommand \taubf{\mbox{\boldmath$\tau$\unboldmath}}

\newcommand \varphibf{\mbox{\boldmath$\varphi$\unboldmath}}

\newcommand \psibf{\mbox{\boldmath$\psi$\unboldmath}}

\newcommand \Omegabf{\hbox{$\bf \Omega$}}

\newcommand \abf{{\bf a}}
\newcommand \bbf{{\bf b}}
\newcommand \cbf{{\bf c}}
\newcommand \dbf{{\bf d}}

\newcommand \gbf{{\bf g}}
\newcommand \hbf{{\bf h}}

\newcommand \pbf{{\bf p}}

\newcommand \rbf{{\bf r}}
\newcommand \sbf{{\bf s}}

\newcommand \wbf{{\bf w}}
\newcommand \xbf{{\bf x}}
\newcommand \ybf{{\bf y}}
\newcommand \zbf{{\bf z}}

\newcommand \Abf{{\bf A}}
\newcommand \Bbf{{\bf B}}
\newcommand \Cbf{{\bf C}}
\newcommand \Dbf{{\bf D}}

\newcommand \Fbf{{\bf F}}
\newcommand \Gbf{{\bf G}}
\newcommand \Hbf{{\bf H}}
\newcommand \Ibf{{\bf I}}

\newcommand \Mbf{{\bf M}}
\newcommand \Nbf{{\bf N}}

\newcommand \Pbf{{\bf P}}
\newcommand \Qbf{{\bf Q}}
\newcommand \Rbf{{\bf R}}

\newcommand \Vbf{{\bf V}}
\newcommand \Wbf{{\bf W}}

\newcommand \Ybf{{\bf Y}}
\newcommand \Zbf{{\bf Z}}

\renewcommand \vec{{\mbox{vec}}}

\newcommand{\circlambda}{\mbox{$\Lambda$
             \kern-.85em\raise1.5ex
             \hbox{$\scriptstyle{\circ}$}}\,}

\newcommand{\diag}{\mathop{\rm diag}}

\def\Re{\mathop{\rm Re}}
\def\Im{\mathop{\rm Im}}

\newcommand{\g}[1]{\bm #1}

\newcommand\blfootnote[1]{%
  \begingroup
  \renewcommand\thefootnote{}\footnote{#1}%
  \addtocounter{footnote}{-1}%
  \endgroup
}

\begin{document}

\title{Channel Estimation and Projection for RIS-assisted MIMO Using Zadoff-Chu Sequences}


\author{Xuemeng Zhou, \textit{Student Member, IEEE,} Zhiyu Yang, \textit{Student Member, IEEE,}\\
Tianyi Zhang, 
Yi Jiang, \textit{Member, IEEE}
}


%


\maketitle
\blfootnote{
The work was supported by National Natural Science Foundation of China Grant No. 61771005. Partial material in this paper appeared in the 10th IEEE/CIC International Conference on Communications in China (ICCC) in Xiamen, China, 28-30 July, 2021. ({\it Corresponding author: Yi Jiang})

The authors are with Key Laboratory for Information Science of Electromagnetic Waves (MoE), Department of Communication Science and Engineering, School of Information Science and Technology, Fudan University, Shanghai, China. (email: yijiang@fudan.edu.cn)

}
\begin{abstract}
The reconfigurable intelligent surface (RIS) technology is a promising enabler for millimeter wave (mmWave) wireless communications, as it can potentially provide  spectral efficiency comparable to the conventional massive multiple-input multiple-output (MIMO) but with significantly lower hardware complexity.
In this paper, we focus on the estimation and projection of the uplink RIS-aided  massive MIMO channel, which can be time-varying. We propose to let the user equipments (UE) transmit Zadoff-Chu (ZC) sequences and let the base station (BS) conduct maximum likelihood (ML) estimation of the uplink channel. The proposed scheme is computationally efficient: it uses ZC sequences to decouple the estimation of the frequency and time offsets; it uses the space-alternating generalized expectation-maximization (SAGE) method to reduce the high-dimensional problem due to the multipaths to multiple lower-dimensional ones per path. Owing to the estimation of the Doppler frequency offsets, the time-varying channel state can be projected, which can significantly lower the overhead of the pilots for channel estimation. The numerical simulations verify the effectiveness of the proposed scheme.

\end{abstract}

\IEEEpeerreviewmaketitle
\section{Introduction}\label{introduction}

Although the millimeter wave (mmWave) communication technology can provide abundant frequency bandwidth, it suffers from large pathloss. The massive multi-input multi-output (MMO) can be used to compensate for the pathloss with its large array gain; thus, mmWave massive MIMO has been intensively researched in the past several years (see \cite{mmMIMO1}, \cite{mmMIMO2}, \cite{mmMIMO3} and the references therein). More recently, the reconfigurable intelligent surface (RIS) technology \cite{RIS_metamaterial} has been introduced to wireless communication systems \cite{RIS}. Compared with the conventional massive MIMO technology, the RIS-assisted MIMO allows for incorporating a very large number (thousands or even more) of reflection elements, leading to the so-called extreme massive MIMO with very large array gain \cite{extremeMIMO}. As a passive device, the RIS can customize favorable wireless propagation environments with limited power consumption. Hence, it is envisioned that the combination of the mmWave and RIS technologies will be one of the keys to the beyond fifth-generation (B5G) wireless communications \cite{RIS_mmWave}.

 It is usually a prerequisite to have the channel state information (CSI) for reaping the great gain of the RIS-assisted massive MIMO. But the CSI  is challenging to estimate due to the high dimensionality of the RIS. The similar issue also occurs in the mmWave phase shifter network (PSN) based hybrid massive MIMO scenario \cite{CE_sparse}, \cite{Fazal2020Channel}. In \cite{onOff1}, Mishra {\it et al.} proposed a least square (LS) based channel estimation method for RIS-aided MIMO. A method using sparse matrix factorization and matrix completion was presented in \cite{Cascaded2020}. However, both methods assumed a frequency-flat channel, which may not be the case in a real-world environment.
 Using the sparsity of the RIS channel, Wan, {\it et al.} proposed a compressive sensing (CS) based algorithm to estimate the frequency-selective channel \cite{wide_Broadband}, but the CS-based broadband channel estimation did not appear robust in different simulation scenarios. In \cite{wide_OFDM}, Zheng {\it et al.} also considered the  scenario of the wide-band and frequency-selective channels.
Assuming high channel correlation between the adjacent elements of the RIS, they proposed to group the adjacent elements into some sub-surfaces to reduce the dimension of the channel estimation problem \cite{wide_OFDM}. But such elements-grouping would cause performance degradation, because the reflection coefficients of all the elements in one group have to be set the same \cite{wide_OFDM}. The channel estimation based on the RIS-element-grouping was extended to the multi-user scenario in \cite{OFDMA} and the time-varying channel scenario \cite{CE_Doppler}. Both \cite{wide_OFDM} and \cite{OFDMA} made the restrictive assumption that the access point is only equipped with a single antenna. {All the algorithms proposed in \cite{wide_OFDM}, \cite{OFDMA} and \cite{CE_Doppler} require that the number of training sequences increased linearly with the number of the elements (or the subsurface) of the RIS, which may cause too much pilot overhead in practice. In \cite{CE_Kalman}, Mao {\it et al.} considered the estimation of a time-variant frequency-flat channels, and employed Kalman filtering to estimate the time-varying channel. 


In this paper, we investigate the uplink channel estimation for the RIS aided massive MIMO-OFDM system in a multipath environment. We consider the scenario where the RIS is placed close to the antennas of the base station (BS); thus, the RIS-to-BS channel can be assumed to be line-of-sight (LOS), frequency-flat, static, and hence is estimated a priori. This assumption is realistic and was also made in \cite{wide_Broadband,zte}. Therefore, we focus on the estimation of the UE-to-RIS channel and the UE-to-BS (direct link) channel, which can be frequency selective and fast time-varying. In contrast, all the aforementioned papers only considered static RIS channels.

We propose to parameterize the multipath channel by a set of the directions of arrival (DOA), the time delays, the channel gains, and the Doppler frequency offsets. As a prominent feature of this work, introducing the Doppler frequency offsets into the model enables projection of the time-varying channel and henceforth reduces the pilots needed for channel estimation, which can help solve the pilot contamination issue \cite{PilotContamination}.


We propose to use Zadoff-Chu (ZC) sequences as the pilots and exploit ZC's unique property of time-delay and frequency offset ambiguity: That is, a ZC sequence with a time delay appears like one with frequency offset. Indeed, this property was exploited in \cite{yj2017zc} to simplify the two-dimensional problem of joint time delay-frequency offset estimation into a problem that can be efficiently solved via two one-dimensional fast Fourier transforms (FFT), based on a conjugate pair of ZC sequences. The ZC sequences' time delay and frequency offset ambiguity is also exploited in this paper to reduce the computational complexity of channel estimation.

We let the UE transmit the ZC sequence multiple times, during which the reflection phases of the RIS will be varied. Based on the multiple sets of received samples, the channel parameters can be estimated. By exploiting the ZC sequence's time delay-frequency offset ambiguity, we devise an FFT-based fast algorithm for joint estimation of the channel parameters. To tackle the multipath scenario, we use the space-alternating generalized expectation-maximization (SAGE) method \cite{SAGE} to decompose the multipath problem into multiple single-path subproblems. For each single path, the ML estimation consists of two steps: i) a fast initialization of the parameter estimates using FFTs; ii) the refinement for super-resolution estimation using Newton's iterative method.

The main contributions of this paper are summarized as follows.
\begin{itemize}
\item[i)] we propose to use ZC sequences as the pilots, and devise a computationally efficient FFT-based channel estimation algorithm by exploiting the ambiguity of the time delay-frequency offset of the ZC sequences;

\item[ii)] the proposed algorithm can achieve super-resolution  maximum likelihood (ML) parameters estimation for the frequency selective multipath channel, which leads to  channel estimation with the root mean square error (RMSE) performance approaching the Cramer-Rao bound (CRB);

\item[iii)] with the estimated Doppler frequency offsets, the time-varying channel state can be projected to greatly reduce the pilot overhead for channel estimation, and thus help mitigate the pilot contamination in massive MIMO communications.
\end{itemize}


The remainder of this paper is organized as follows. Section $\textrm{II}$  establishes the signal model and formulates the channel estimation problem based on RIS. Section $\textrm{III}$ derives the solution to the channel estimation problem in the single-path case when the UE-BS channel is blocked. Based on the single-path solution, we further propose to apply the SAGE method to the multipath case. Section $\textrm{IV}$ addresses the more general case where the direct channel link between the UE and the BS exists. Numerical examples are given in Section $\textrm{V}$ and conclusions are made in Section $\textrm{VI}$.

Notations: $(\cdot)^T$, $(\cdot)^*$ and $(\cdot)^H$ stand for transpose, conjugate and Hermitian transpose, respectively. $\otimes$  denotes Kronecker product and $\odot$ denotes Hadamard (element-wise) product. ${\mathbb Z}$ is the set of integers, ${\mathbb R}$ is the set of real numbers, and $\mathbb{C}^{N\times K}$ is the set of $N\times K$ complex matrices. $\diag(\abf)$ denotes a diagonal matrix with vector $\abf$ being its diagonal and $\vec(\cdot)$ denotes a vectorization operation to a matrix by stacking the columns of the matrix into a long column-vector. $\left\arrowvert\cdot\right\arrowvert$ stand for absolute value, $\|\cdot\|_F$ stands for the Frobenius norm, and $\left\Arrowvert\cdot\right\Arrowvert$ the $l_2$ norm. $\Re\{\cdot\}$ and $\Im\{\cdot\}$ stands for taking the real and imaginary part, respectively. $[\cdot]_{i,j}$ denotes the $(i,j)$th element of a matrix.

\section{Signal Model and Problem Formulation}
\subsection{Signal Model}

\begin{figure}[ht]
\centering
{\psfig{figure=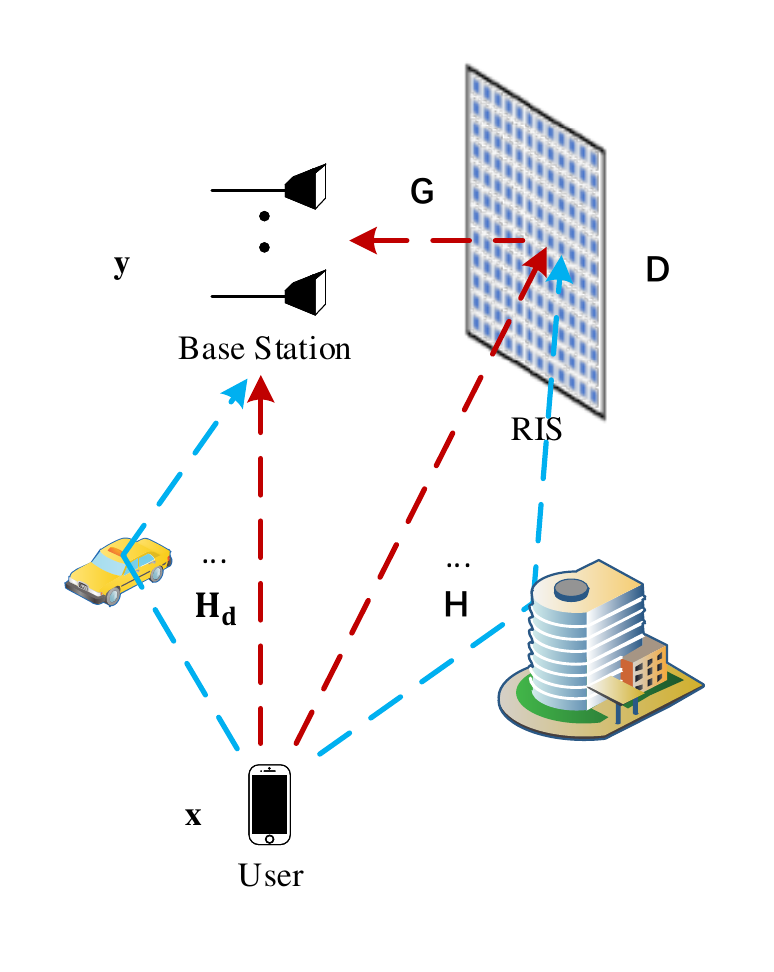 ,width=2.5 in}}
\caption{A RIS aided massive MIMO wireless communication system.}\label{fig:RIS}
\end{figure}
An RIS aided massive MIMO wireless communication system consists of a BS, a RIS, and a UE, as shown in Fig. \ref{fig:RIS}, where the BS is connected to the UE through the direct link and the reflections of the RIS. Given that the RIS is a planar uniform rectangular array (URA) of $P$ rows and $Q$ columns. The array response of the RIS with respect to a signal from angle $(\theta,\varphi)$ can be represented by $\Abf(\theta,\varphi)\in {\mathbb C}^{P\times Q}$, whose entries are \cite{DOA}
\begin{align}\label{eq:abf}
\left[\Abf(\theta,\varphi)\right]_{pq}=e^{-j\frac{2\pi}{\lambda}(q-1)d_x\sin\varphi\cos\theta+j\frac{2\pi}{\lambda}(p-1)d_z\cos\varphi}, \nonumber\\
p = 1,2,\cdots,P, q = 1,2,\cdots,Q,
\end{align}
where $\lambda$ denotes the wavelength of the carrier frequency, $d_x$ and $d_z$ represent the inter-element spacing along the horizontal and vertical direction of the RIS, respectively.

The array response of the $M_r$-antenna BS with respect to a signal from angle $\overline{\theta}$ can be represented by
\ben\label{eq:cbf}
\left[\cbf(\overline{\theta})\right]_{m}=e^{-j\frac{2\pi}{\lambda}(m-1)\bar{d}_x\cos\overline{\theta}} \in {\mathbb C}^{M_r\times 1}, m = 1,2,\cdots,M_r,
\een
where $\bar{d}_x$ is the inter-antenna distance of the BS array.

In the uplink channel, while the single-antenna UE transmits a continuous-time signal $x(t)$, the BS receives the signal
\begin{align}\label{eq:yt}
\ybf(t)&=\Gbf\Dbf\int \Hbf(\tau,t) x(t-\tau) d\tau\nonumber\\
&+\int \Hbf_d(\tau,t) x(t-\tau) d\tau+\zbf(t),
\end{align}
where $\Gbf \in {\mathbb C}^{M_r \times M}$ is the RIS-to-BS channel with $M=PQ$ being the number of the elements of the RIS, $\Dbf= \diag \{e^{j\phi_1},\ldots, e^{j\phi_M} \}$ with the diagonal represents the reflection phases of the RIS elements, $\Hbf(t,\tau)$ is the time-varying multipath channel between the UE and the RIS, and $\Hbf_d(t,\tau)$ is the time-varying multipath channel between the UE and the BS, $\zbf(t)\sim {\cal CN}(0,\sigma_z^2\Ibf_{M_r})$ is the additive white Gaussian noise. Throughout this paper, we assume that the antennas of the BS are placed close to the RIS so that the channel $\Gbf$ is frequency flat and static. Indeed, we can expect a steady LOS-path between the RIS and the BS, while the NLOS paths are much weaker than the LOS paths in the mmWave frequency band as shown in \cite{similar_G}. Therefore, one can construct the LOS part based on the position of the RIS relative to the BS. The same assumption that $\Gbf$ is known a priori is also made in \cite{wide_Broadband}. This paper focuses on estimating the UE-to-RIS channel
\begin{align}\label{eq:H}
\Hbf(\tau,t)=&\sum_{u=1}^{U}\hbf_u(t) \delta(\tau-\tau_u),
\end{align}
and the UE-to-BS channel
\begin{align}\label{eq:Hd}
\Hbf_d(\tau,t)=&\sum_{d=1}^{D}\hbf_d(t) \delta(\tau-\overline{\tau}_d).
\end{align}
Inserting (\ref{eq:H}) and (\ref{eq:Hd}) into (\ref{eq:yt}) yields
\ben\label{eq:yt2}
\ybf(t)=\Gbf\Dbf\sum_{u=1}^{U}\hbf_u(t) x(t-\tau_u)+\sum_{d=1}^{D}\hbf_d(t) x(\tau-\overline{\tau}_d)+\zbf(t).
\een

As the Doppler frequency spread is proportional to $\frac{v}{C}f_c$ -- where $v$ is the UE's mobility speed, $C$ is the speed-of-light, and $f_c$ is the carrier frequency -- the large $f_c$ of the mmWave makes it necessary to model the channel as time-varying. To estimate the time-varying and high-dimensional channels $\Hbf(\tau,t)$ and $\Hbf_d(\tau,t)$ will apparently entail a large overhead of training sequences. To reduce the overhead, we propose to parameterize the time-varying channel as follows.

Denote $\abf(\theta_u,\varphi_u)\triangleq \vec(\Abf(\theta_u,\varphi_u))\in {\mathbb C}^{M}$, the array response of the $u$th path in (\ref{eq:yt2}) can be represented as
\ben
\hbf_u(t) = \beta_u\abf(\theta_u,\varphi_u) e^{j 2\pi \xi_u t},
\een
where $\beta_u$ is the complex-valued multipath gain, and $\xi_u$ is the Doppler frequency offset.
The array response of the $d$th path in (\ref{eq:yt2}) can be represented as
\ben
\hbf_d(t) = \alpha_d\cbf(\overline{\theta}_u) e^{j 2\pi \overline{\xi}_d t},
\een
where $\alpha_d$ is the complex-valued multipath gain, and $\overline{\xi}_d$ is the Doppler frequency offset.
Thus, we rewrite (\ref{eq:yt2}) to be
\begin{align}
\ybf(t)=\Gbf\Dbf\sum_{u=1}^{U}\beta_u \abf(\theta_u,\varphi_u) x(t-\tau_u)e^{j2\pi\xi_u t} \nonumber\\
+\sum_{d=1}^{D}\alpha_d \cbf(\overline{\theta}_d) x(t-\overline{\tau}_d)e^{j2\pi\overline{\xi}_d t}+\zbf(t).
\end{align}

The sampled  signal at the output of the receiver's analog-to-digital converters (ADC) is
\begin{align} \label{eq.ySamle1}
\ybf(n)=\Gbf\Dbf\sum_{u=1}^{U}\beta_u\abf(\theta_u,\varphi_u) x(n-\tau_u)e^{j2\pi\xi_u n}  \nonumber\\
+\sum_{d=1}^{D}\alpha_d \cbf(\overline{\theta}_d) x(n-\overline{\tau}_d)e^{j2\pi\overline{\xi}_d n}+\zbf(n), n\in {\mathbb Z},
\end{align}
where we denote without loss of generality the Nyquist sampling interval $T_s=1$ for notational simplicity, but $\tau_u \in {\mathbb R}$ and $\overline{\tau}_d \in {\mathbb R}$ are not necessarily integers.

To improve the channel estimation, we change the reflection phases of the RIS for $K$ times, each corresponding to the transmission of a pilot with length $L$. Given that the channel parameters are static during the $K$ observations (note that a time-varying channel can be represented with static parameters), the BS receives
\begin{align}\label{eq.ySamle2}
\ybf_k(n)&=
\Wbf(\Phi_k)\sum_{u=1}^{U}\beta_u e^{j2\pi\xi_u (k-1)N}\abf(\theta_u,\varphi_u)x(n-\tau_u)\nonumber\\
&\times e^{j2\pi\xi_u n}+\sum_{d=1}^{D}\alpha_d e^{j2\pi\overline{\xi}_d (k-1)N}\cbf(\overline{\theta}_d)x(n-\overline{\tau}_d)\nonumber\\
&\times e^{j2\pi\overline{\xi}_d n}+\zbf_k(n), k=1,2,\cdots,K,
\end{align}
where $\Wbf(\Phi_k)\triangleq\Gbf\Dbf(\Phi_k)$ with $\Phi_k$ being the phase of all elements of the RIS for the $k$th observation, and $N$ is the length of the OFDM symbol.

For each observation, we assume that $L$ samples, with indices from $-L/2$ to $L/2-1$, are processed [if $L$ is an odd number, the indices are from $-(L-1)/2$ to $(L-1)/2$]. The index range differs from the convention to cater to the proposed special design of the pilot $x(t)$, as we will see soon.

By formatting $\ybf_k(n)$ into the matrices
\ben
\begin{split}
&\Ybf_k \!=\! [\ybf_k(-\frac{L}{2}),\ybf_k(-\frac{L}{2}+1),\dots, \ybf_k(\frac{L}{2}-1)]\!\in\mathbb{C}^{M_r\times L}, \\
\mbox{and}\\
&\Zbf_k=\left[\zbf_k(-L/2),\zbf_k(-L/2+1),\cdots,\zbf_k(L/2-1)\right],
\end{split}
\een
we have from (\ref{eq.ySamle2}) that
\begin{align}\label{eq.yMatix}
\Ybf_k \!=\! \Wbf(\Phi_k)\sum_{u=1}^{U}\abf(\theta_u,\varphi_u)\beta_ue^{j2\pi\xi_u (k-1)N}\left(\xbf(\tau_u)\odot \dbf(\xi_u)\right)^T\nonumber\\
\!+\!\sum_{d=1}^{D}\cbf(\overline{\theta}_d)\alpha_d e^{j2\pi\overline{\xi}_d (k-1)N}\left(\xbf(\overline{\tau}_d)\odot \dbf(\overline{\xi}_d)\right)^T\!+\!\Zbf_k,
\end{align}
where
\ben\label{eq.xtauxi}
\begin{split}
\xbf(\tau)=\begin{bmatrix}
x(-\frac{L}{2}-\tau) \\ x(-\frac{L}{2}+1-\tau) \\
\vdots\\
x(\frac{L}{2}-1-\tau)
\end{bmatrix},
\; \dbf(\xi) = \begin{bmatrix}e^{j2\pi\xi(-\frac{L}{2})}\\e^{j2\pi\xi(-\frac{L}{2}+1)}\\
\vdots\\
e^{j2\pi\xi (\frac{L}{2}-1)}
\end{bmatrix}.
\end{split}
\een

Denoting
\ben\label{mathW}
{\mathcal{W}}(\xi_u)\triangleq\begin{bmatrix}\Wbf(\Phi_1) \\ \Wbf(\Phi_2)e^{j2\pi\xi_u N} \\ \vdots \\ \Wbf(\Phi_{K})e^{j2\pi\xi_u(K-1)N} \end{bmatrix}\in{\mathbb C}^{KM_r \times M}
\een
and
\ben\label{BMathcal}
{\mathcal{B}}(\overline{\xi}_d)\triangleq\begin{bmatrix}\Ibf_{M_r} \\ \Ibf_{M_r}e^{j2\pi\overline{\xi}_d N} \\ \vdots \\ \Ibf_{M_r}e^{j2\pi\overline{\xi}_d(K-1)N} \end{bmatrix}
=\pbf(\overline{\xi}_d) \otimes \Ibf_{M_r}
\in{\mathbb C}^{KM_r \times M_r}
\een
where
\ben \label{eq.pxi}
\pbf(\overline{\xi}_d) = \begin{bmatrix}1 \\ e^{j2\pi\overline{\xi}_d N} \\ \vdots \\ e^{j2\pi\overline{\xi}_d (K-1)N}\end{bmatrix}\in{\mathbb C}^{K \times 1},
\een
we stack the received samples as
\begin{align}\label{eq.Y}
\Ybf = \sum_{u=1}^{U}{\cal W}(\xi_u)\abf(\theta_u,\varphi_u)\beta_u(\xbf(\tau_u)\odot \dbf(\xi_u))^T\nonumber\\
+\sum_{d=1}^{D}{\cal B}(\overline{\xi}_d)\cbf(\overline{\theta}_d)\alpha_d(\xbf(\overline{\tau}_d)\odot \dbf(\overline{\xi}_d))^T+\Zbf,
\end{align}
where
\ben
\begin{split}
&\Ybf=\begin{bmatrix} \Ybf_1^T, \Ybf_2^T,\cdots , \Ybf_K^T\end{bmatrix}^T \in {\mathbb C}^{KM_r\times L}, \\
&\Zbf=\begin{bmatrix} \Zbf_1^T, \Zbf_2^T,\cdots , \Zbf_K^T\end{bmatrix}^T \in {\mathbb C}^{KM_r\times L}.
\end{split}
\een

\subsection{Problem Formulation}

Given the whiten Gaussian noise, the ML estimation of the channel parameters is identical to the least squared one:
\begin{align}\label{eq.MLOri}
\{\hat{\betabf},\hat{\taubf},\hat{\xibf},\hat{\thetabf},\hat{\varphibf},\hat{\alphabf},\hat{\overline{\taubf}},\hat{\overline{\xibf}},\hat{\overline{\thetabf}}\}=\arg\min_{\betabf,\taubf,\xibf,\thetabf,\varphibf,\alphabf,\overline{\taubf},\overline{\xibf},\overline{\thetabf}}\nonumber\\
\Arrowvert \Ybf - \sum_{u=1}^{U}{\cal W}(\xi_u)\abf(\theta_u,\varphi_u)\beta_u(\xbf(\tau_u)\odot \dbf(\xi_u))^T \nonumber\\
 - \sum_{d=1}^{D}{\cal B}(\overline{\xi}_d)\cbf(\overline{\theta}_d)\alpha_d(\xbf(\overline{\tau}_d)\odot \dbf(\overline{\xi}_d))^T \Arrowvert^2_F,
\end{align}
where $\betabf=[\beta_1,\cdots,\beta_U]^T,\taubf=[\tau_1,\cdots,\tau_U]^T,\xibf=[\xi_1,\cdots,\xi_U]^T,\thetabf=[\theta_1,\cdots,\theta_U]^T,\varphibf=[\varphi_1,\cdots,\varphi_U]^T,
\alphabf=[\alpha_1,\cdots,\alpha_D]^T,\overline{\taubf}=[{\overline{\tau}}_1,\cdots,{\overline{\tau}}_D]^T,\overline{\xibf}=[{\overline{\xi}}_1,\cdots,{\overline{\xi}}_D]^T,\overline{\thetabf}=[\overline{\theta}_1,\cdots,\overline{\theta}_D]^T$.

The high-dimensional problem (\ref{eq.MLOri}) appears highly involved. We propose to use ZC sequences as the pilot $x(t)$, which will be shown to be able to drastically simplify this problem.

A length-$\tilde{L}$ ZC sequence is\cite{chu1972polyphase}
\ben\label{eq.zcdefin}
s(n)=\begin{cases}e^{\frac{j\pi rn(n+1)}{\tilde{L}}}&{\rm if}\;\tilde{L}\;{\rm is\;odd} \cr e^{\frac{j\pi rn^2}{\tilde{L}}}&{\rm if}\;\tilde{L}\;{\rm is\;even,}\end{cases}
\een
where the index $r$ is an integer co-prime to $\tilde{L}$.

It is easy to verify that $s(n)=s(n+\tilde{L})$, i.e., the ZC is periodic. Hence we can set the index range of the ZC to be from $-\tilde{L}/2$ to $\tilde{L}/2-1$ for an even $\tilde{L}$, or from $-(\tilde{L}-1)/2$ to $(\tilde{L}-1)/2$ for an odd $\tilde{L}$.
\begin{figure}[ht]
\centering
{\psfig{figure=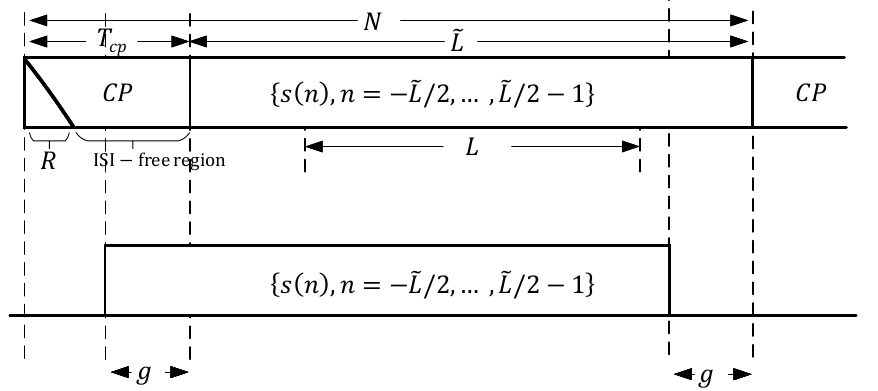 ,width=3.2in}}
\caption{Illustration of the truncation process and relative delay between the transmitted signal and the $\tilde{L}$ samples.}\label{fig:sendSig1}
\end{figure}
Selecting the ZC sequence\footnote{Here and in remainder of this paper, we only consider the ZCs with index $r=1$.} as follows:
\ben\label{eq.sn11}
s(n)=e^{j\pi n^2/\tilde{L}},n=-\frac{\tilde{L}}{2},-\frac{\tilde{L}}{2}+1,\cdots,\frac{\tilde{L}}{2}-1,
\een
we can add a length-$T_{cp}$ cyclic prefix (CP)
\ben  \nonumber
\begin{split}
&\left[ s\left(\frac{\tilde{L}}{2}-T_{cp}\right), s\left(\frac{\tilde{L}}{2}-T_{cp}+1\right),\cdots, s\left(\frac{\tilde{L}}{2}-1\right)\right]
\end{split}
\een
to transform a  \textit{$linear$} time delay to a \textit{$cyclic$} delay.

While the high-frequency components of the ZC sequence will be affected by the pulse shaping filter, the low-frequency components will be intact. Indeed, it is established in  \cite[eq. (22)]{JEVAR} that the lower-frequency part of the ZC sequence pulse shaped by a raised cosine filter can be represented as a continuous-time chirp signal $x(t)=e^{j\pi t^2/\tilde{L}}$ for $-\frac{L}{2}\le t< \frac{L}{2}$, which has the ambiguity between time-delay and frequency-offset:
\begin{align}\label{eq.tdfo}
x(t-\tau)
=e^{j\frac{\pi \tau^2}{\tilde{L}}}e^{-j\frac{2\pi\tau t}{\tilde{L}}}x(t),\;  -\frac{L}{2}\le t< \frac{L}{2};
\end{align}
that is, a time delay $\tau$ amounts to a frequency offset $\frac{\tau}{\tilde{L}}$.

Denoting \ben
\xbf(\tau) \triangleq \begin{bmatrix}
x(-\frac{L}2-\tau)\\
x(-\frac{L}2+1-\tau)\\
\vdots\\
x(\frac{L}2-1-\tau)
\end{bmatrix},\; \sbf  \triangleq \begin{bmatrix}
s(-\frac{L}2)\\
s(-\frac{L}2+1)\\
\vdots\\
s(\frac{L}2-1)
\end{bmatrix},
\een
and recognizing that $\xbf(0) = \sbf$,
we can rewrite (\ref{eq.tdfo}) in the vector form as
\ben \label{x-tau}
\xbf(\tau) = e^{j\frac{\pi \tau^2}{\tilde{L}}} \sbf \odot \dbf\left(-\frac{\tau}{\tilde{L}}\right),
\een
where $\dbf(\cdot)$ is as defined in (\ref{eq.xtauxi}).

To have the time-frequency transformation relationship (\ref{eq.tdfo}) hold, we partially remove the CP,
making sure $\ybf(0)$ is sampled in the ``ISI-free region'' (illustrated in Fig. \ref{fig:sendSig1}).
That is, the delayed version of the transmitted signal should be sampled from the ``ISI-free region'', otherwise it may suffer from interferences outside of the sequence.
In Fig. \ref{fig:sendSig1}, $\ybf(0)$ is sampled $g$-sample ahead the time of $s(0)$,
where $T_{cp}-R>g\geq 0$ with the assumption that all the propagation delays are smaller than $R$, which is known a prior.
Although the pilot is of length $N=\tilde{L}+ T_{cp}$ samples, we only truncate out $\tilde{L}$ samples, indexed from $-\tilde{L}/2$ to $\tilde{L}/2-1$, starting from the ISI-free region as shown in Fig. \ref{fig:sendSig1}, out of which we further truncate out the low frequency part of the $L$ samples indexed from $-L/2$ to $L/2-1$ for channel estimation. Note again that $L<\tilde{L}$ is chosen to ensure that the ZC sequence filtered by a raised cosine filter is a chirp signal that satisfy (\ref{eq.tdfo}).

Inserting (\ref{x-tau}) into (\ref{eq.MLOri}), yields
\begin{align}\label{eq.MLOri2}
\{\hat{\betabf},\hat{\taubf},\hat{\xibf},\hat{\thetabf},\hat{\varphibf},\hat{\alphabf},\hat{\overline{\taubf}},\hat{\overline{\xibf}},\hat{\overline{\thetabf}}\}=\arg\min_{\betabf,\taubf,\xibf,\thetabf,\varphibf,\alphabf,\overline{\taubf},\overline{\xibf},\overline{\thetabf}}\nonumber\\
\Arrowvert \Ybf - \sum_{u=1}^{U}{\cal W}(\xi_u)\abf(\theta_u,\varphi_u)\beta_u e^{j\frac{\pi \tau_u^2}{\tilde{L}}} \dbf\left(\xi_u-\frac{\tau_u}{\tilde{L}}\right) ^T\diag(\sbf) \nonumber\\
-\sum_{d=1}^{D}{\cal B}(\overline{\xi}_d)\cbf(\overline{\theta}_d)\alpha_d e^{j\frac{\pi \overline{\tau}_d^2}{\tilde{L}}} \dbf\left(\overline{\xi}_d-\frac{\overline{\tau}_d}{\tilde{L}}\right) ^T\diag(\sbf) \Arrowvert^2_F.
\end{align}
Denote
\ben\label{betaT}
\tilde{\beta}_u \triangleq \beta_u e^{\frac{j\pi \tau_u^2}{\tilde{L}}},\;\tilde{\alpha}_d \triangleq \alpha_d e^{\frac{j\pi \overline{\tau}_d^2}{\tilde{L}}}
\een
and
\ben\label{eq.zeta}
\zeta_u\triangleq\xi_u-\frac{\tau_u}{\tilde{L}},\;\overline{\zeta}_d\triangleq\overline{\xi}_d-\frac{\overline{\tau}_d}{\tilde{L}}.
\een
Then (\ref{eq.MLOri2}) can be expressed as
\begin{align}\label{eq.MLOri3}
\{\hat{\tilde{\betabf}},\hat{\zetabf},\hat{\xibf},\hat{\thetabf},\hat{\varphibf},\hat{\tilde{\alphabf}},\hat{\overline{\zetabf}},\hat{\overline{\xibf}},\hat{\overline{\thetabf}}\}=\arg\min_{\tilde{\betabf},\zetabf,\xibf,\thetabf,\varphibf,\tilde{\alphabf},\overline{\zetabf},\overline{\xibf},\overline{\thetabf}}\nonumber\\
\Arrowvert \Ybf - \sum_{u=1}^{U}{\cal W}(\xi_u)\abf(\theta_u,\varphi_u)\tilde{\beta}_u\dbf(\zeta_u) ^T\diag(\sbf) \nonumber\\
-\sum_{d=1}^{D}{\cal B}(\overline{\xi}_d)\cbf(\overline{\theta}_d)\tilde{\alpha}_d\dbf(\overline{\zeta}_d) ^T\diag(\sbf)\Arrowvert^2_F.
\end{align}
Denote
\ben \label{eqYtilde} \tilde{\Ybf}\triangleq\Ybf\diag(\sbf^*) , \; {\rm and} \; \tilde{\Ybf}=[\tilde{\Ybf}_1^T,\cdots,\tilde{\Ybf}_K^T]^T,\een
then we can rewrite (\ref{eq.MLOri3}) to be
\begin{align}\label{eq.MLOri4}
\{\hat{\tilde{\betabf}},\hat{\zetabf},\hat{\xibf},\hat{\thetabf},\hat{\varphibf},\hat{\tilde{\alphabf}},\hat{\overline{\zetabf}},\hat{\overline{\xibf}},\hat{\overline{\thetabf}}\}=\arg\min_{\tilde{\betabf},\zetabf,\xibf,\thetabf,\varphibf,\tilde{\alphabf},\overline{\zetabf},\overline{\xibf},\overline{\thetabf}}\nonumber\\
\Arrowvert \tilde{\Ybf} - \sum_{u=1}^{U}{\cal W}(\xi_u)\abf(\theta_u,\varphi_u)\tilde{\beta}_u\dbf(\zeta_u) ^T\nonumber\\
- \sum_{d=1}^{D}{\cal B}(\overline{\xi}_d)\cbf(\overline{\theta}_d)\tilde{\alpha}_d\dbf(\overline{\zeta}_d) ^T \Arrowvert^2_F.
\end{align}
The next two sections are dedicated to solve this high-dimensional nonlinear problem, first without the direct link between the UE and the BS and then with the link. After solving (\ref{eq.MLOri4}), we have from (\ref{eq.zeta}) the time delays as
\begin{align}\label{eq.interPoint}
\hat{\tau}_u=(\hat\xi_u-\hat\zeta_u)\tilde{L},\;\hat{\overline{\tau}}_d=(\hat{\overline{\xi}}_d-\hat{\overline{\zeta}}_d)\tilde{L} \nonumber\\
u=1,\cdots,U, d=1,\cdots,D.
\end{align}


\section{Estimation of the RIS Channel Without the UE-to-BS Direct Link}
In this section, we study the channel estimation problem in the  scenario where the direct link between the UE and the BS is blocked. The math derivations in this simpler case will lay foundation for the channel estimation algorithm developed in Section \ref{sec4} for the more general case with the direct link.

Assuming no direct link between the UE and the BS, we can rewrite (\ref{eq.ySamle1}) as
\begin{align}
\ybf(n)=\Gbf\Dbf\sum_{u=1}^{U}\beta_u\abf(\theta_u,\varphi_u) x(n-\tau_u)e^{j2\pi\xi_u n}+\zbf(n),
\end{align}
and (\ref{eq.MLOri4}) can be simplified as
\begin{align}\label{eq.MLOri5}
&\{\hat{\tilde{\betabf}},\hat{\zetabf},\hat{\xibf},\hat{\thetabf},\hat{\varphibf}\}=\arg\min_{\tilde{\betabf},\zetabf,\xibf,\thetabf,\varphibf}\nonumber\\
&\left\Arrowvert \tilde{\Ybf} - \sum_{u=1}^{U}{\cal W}(\xi_u)\abf(\theta_u,\varphi_u)\tilde{\beta}_u\dbf(\zeta_u) ^T\right\Arrowvert^2_F.
\end{align}
For ease of presentation, we first address the single path case.

\subsection{The Channel Estimation in Single-path Case}\label{subsection:H}
In the single-path case, we can rewrite (\ref{eq.MLOri5}) to be
\ben\begin{split}\label{eq.ML}
&\{\hat{\tilde{\beta}},\hat{\zeta},\hat{\xi},\hat{\theta},\hat{\varphi}\}\\
&=\arg\min_{\tilde{\beta},\zeta,\xi,\theta,\varphi}\left\Arrowvert \tilde{\Ybf} - \tilde{\beta} {\cal W}(\xi)\abf(\theta,\varphi)\dbf(\zeta) ^T\right\Arrowvert^2_F.
\end{split}\een
Expanding  (\ref{eq.ML}) to obtain the quadratic function of $\tilde{\beta}$, we can minimize the function with respect to $\tilde{\beta}$ to obtain
\ben\label{eq.betaTilde}
\tilde{\beta}=\frac{\abf(\theta,\varphi)^H{\cal W}(\xi)^H\tilde{\Ybf}\dbf(\zeta)^*}{L\times\abf(\theta,\varphi)^H{\cal W}(\xi)^H{\cal W}(\xi)\abf(\theta,\varphi)}.
\een
It follows from  (\ref{mathW}) that (\ref{eq.betaTilde}) can be rewritten as
\begin{align}\label{eq.betaTildeNew}
&\tilde{\beta}=\frac{\abf(\theta,\varphi)^H{\cal W}(\xi)^H\tilde{\Ybf}\dbf(-\zeta)}
{L\times\sum_{k=1}^{K}\left\Arrowvert\abf(\theta,\varphi)^H\Wbf(\Phi_k)^H\right\Arrowvert^2}.
\end{align}

Inserting (\ref{eq.betaTildeNew}) into (\ref{eq.ML}) yields
\begin{align}
&\{\hat{\zeta},\hat{\xi},\hat{\theta},\hat{\varphi}\}
=\arg\max_{\zeta,\xi,\theta,\varphi} \frac{\left| \abf(\theta,\varphi)^H{\cal W}(\xi)^H\tilde{\Ybf}\dbf(-\zeta) \right|^2}{\sum_{k=1}^{K}\left\Arrowvert\abf(\theta,\varphi)^H\Wbf(\Phi_k)^H\right\Arrowvert^2} \label{eq.MLnew1}
 \\
&\!=\!\arg\!\max_{\zeta,\xi,\theta,\varphi} \!\!\!\frac{\left| \sum_{k=1}^{K}\!\abf(\theta,\varphi)^H\Wbf(\Phi_k)^H\tilde{\Ybf}_k\dbf(-\zeta)e^{-j2\pi N\xi (k-1)} \right|^2}{\sum_{k=1}^{K}\left\Arrowvert\abf(\theta,\varphi)^H\Wbf(\Phi_k)^H\right\Arrowvert^2} \label{eq.MLnew2},
\end{align}
which can be coarsely solved using FFTs as explained in the following.

\subsubsection{Coarse Solution to (\ref{eq.MLnew2}) Using FFTs} \label{subsec:fft}

The denominator of (\ref{eq.MLnew2}) can be rewritten as
\ben\label{eq.denominator}
{\sum_{k=1}^{K}\sum_{m=1}^{M_r}\left|\abf(\theta,\varphi)^H\wbf_m^{(k)}\right|^2},
\een
where $\wbf_m^{(k)}$ is the $m$th column of $\Wbf(\Phi_k)^H$.
We can get
\ben\begin{split}
&\abf(\theta,\varphi)^H\wbf_m^{(k)}\\
&=\sum\limits_{p=1}^{P}\sum\limits_{q=1}^{Q}w_{q,p,m}^{(k)}
e^{j\frac{2\pi}{\lambda}(q-1)d_x\sin\varphi\cos\theta-j\frac{2\pi}{\lambda}(p-1)d_z\cos\varphi}, \end{split}\een
which can be evaluated by applying 2D-FFT to $\wbf_m^{(k)}$, and $w_{q,p,m}^{(k)}$ is the $((p-1)Q+q)$th element of $\wbf_m^{(k)}$; thus, (\ref{eq.denominator}) can be evaluated efficiently over the mesh grids $-\frac{d_x}{\lambda}\sin\varphi\cos\theta = 0, \frac{1}{N_\theta}, \ldots, \frac{N_\theta-1}{N_\theta} , \frac{d_z}{\lambda}\cos\varphi = 0, \frac{1}{N_\varphi}, \ldots, \frac{N_\varphi-1}{N_\varphi}$ using $N_\theta \times N_\varphi$-point  2D-FFT $KM_r$ times.

Since $\{\Wbf(\Phi_k),k=1,\cdots,K\}$ are prescribed, we can actually compute (\ref{eq.denominator}) offline and store in advance
\ben
\Rbf^{(k)}\!\triangleq\! \begin{bmatrix} \begin{array}{c}\vdots\\ \abf(\theta,\varphi)^H\wbf_1^{(k)} \\
\vdots \end{array} \cdots \begin{array}{c}\vdots\\ \abf(\theta,\varphi)^H\wbf_{M_r}^{(k)} \\
\vdots \end{array} \end{bmatrix}\!\in {\mathbb C}^{N_{\theta}N_{\varphi}\times M_r},
\een
where the $N_{\theta} N_{\varphi}$-dimensional columns correspond to the $N_{\theta}\times N_{\varphi}$ points 2D-FFT of the corresponding columns of $\Wbf(\Phi_k)^H$. Then the denominator of (\ref{eq.MLnew2}) is evaluated and stored in vector $\g{\gamma} \in {\mathbb C}^{N_{\theta}N_{\varphi}}$ whose entries are
 \ben
 \gamma_n = \sum_{k=1}^K \| \Rbf^{(k)}(n,:)\|^2,
 \een
where $\Rbf^{(k)}(n,:)$ stands for the $n$th row of $\Rbf^{(k)}$.

According to the expression $\dbf(\xi_u)$ given in (\ref{eq.xtauxi}), in the numerator of (\ref{eq.MLnew2})
$\tilde{\Ybf}_k\dbf(-\zeta)$ can be evaluated efficiently over the mesh grids $\zeta = 0, \frac{1}{N_\zeta}, \ldots, \frac{N_\zeta-1}{N_\zeta}$ using $N_\zeta$-point FFTs to obtain
 \ben \label{eqBbf}
\Bbf^{(k)}\triangleq\begin{bmatrix}\tilde{\Ybf}_k\dbf(0), \cdots, \tilde{\Ybf}_k\dbf(-\frac{N_\zeta-1}{N_\zeta}) \end{bmatrix}\in {\mathbb C}^{M_r\times N_\zeta}.
\een
Denote
\ben \label{eqQbf}
\Qbf^{(k)}\triangleq\Rbf^{(k)}\Bbf^{(k)}\in {\mathbb C}^{({N_{\theta}N_{\varphi}})\times N_{\zeta}},
\een
which corresponds to the term $\abf(\theta,\varphi)^H\Wbf(\Phi_k)^H\tilde{\Ybf}_k\dbf(-\zeta)$
 in the numerator of (\ref{eq.MLnew2}). Thus, we can evaluate the numerator of (\ref{eq.MLnew2}) by applying $N_{\xi}$-point FFTs across the $K$ matrices $\Qbf^{(k)}$ to obtain a 3D tensor $\Mbf \in {\mathbb C}^{N_{\theta}N_{\varphi}\times  N_\zeta \times N_\xi}$, i.e.,
 \ben \label{eqQfft}  \Mbf(:,:,n_\xi) = \sum_{k=1}^{K}\Qbf^{(k)} e^{-j2\pi \frac{(n_\xi-1)(k-1)}{N_\xi} }, n_\xi=1,2,\ldots, N_\xi.
 \een
Thus, the numerator of (\ref{eq.MLnew2}) can be evaluated by computing the squared absolute values of every entries of $\Mbf$, which we refer to as $\Nbf = |\Mbf|^2$. To evaluate the numerator $\Nbf$ divided by the  denominator $\g{\gamma}$, we compute an $N_{\theta}N_{\varphi}\times  N_\zeta\times N_\xi$ real-valued matrix $\breve{\Abf}$ by
\ben
\breve{\Abf}(n,n_\zeta,n_\xi) =  \Nbf(n,n_\zeta,n_\xi)/\gamma_n.
\een

By locating the largest entry of $\breve{\Abf}$, 
whose subscript is denoted by $(n_\varphi,n_\theta,n_\zeta,n_\xi)$, we calculate
\ben
f_x\triangleq\frac{n_\varphi-1}{N_\varphi},f_y\triangleq\frac{n_\theta-1}{N_\theta},f_\zeta\triangleq\frac{n_\zeta-1}{N_\zeta},f_\xi\triangleq\frac{n_\xi-1}{N_\xi}, \label{eqffff}
\een
of which all range from 0 to 1. We then take the modulo operation
\ben
f = \left\{\begin{array}{ll} f, & f < 0.5\\ f-1, & f\ge 0.5 \end{array} \right.
\een
so that the frequencies in (\ref{eqffff}) range from $-0.5$ to $0.5$. Therefore, the  channel parameters can be (coarsely) estimated as
\ben\label{FFTpara}
\begin{aligned}
&\varphi=\arccos\left(\frac{\lambda f_x}{d_z}\right),\; \theta=\arccos\left(-\frac{\lambda f_y}{d_x\sin\varphi}\right),\\
&\hspace{4em} \zeta=f_\zeta,\quad \xi=f_\xi.
\end{aligned}
\een


The online computation involves (\ref{eqBbf})(\ref{eqQbf}) and (\ref{eqQfft}), which take $\mathcal{O}(KM_rN_{\zeta}\log(N_{\zeta}))$ flops,
$\mathcal{O}(KN_{\theta}N_{\varphi}M_rN_{\zeta})$ flops, and $\mathcal{O}(N_{\theta}N_{\varphi}N_{\zeta}N_{\xi}\log(N_{\xi}))$, respectively.
Hence, the total online computational complexity is $\mathcal{O}(KM_rN_{\zeta}\log(N_{\zeta})+KN_{\theta}N_{\varphi}M_rN_{\zeta}+N_{\theta}N_{\varphi}N_{\zeta}N_{\xi}\log(N_{\xi}))$ flops, a drastically reduced complexity compared with the simplistic method of computing (\ref{eq.MLnew2}) over $N_{\theta}\times N_{\varphi}\times N_{\zeta}\times N_{\xi}$ mesh grid points, which requires $\mathcal{O}(N_{\theta}N_{\varphi}N_{\zeta}N_{\xi}PQLK)$ flops instead.

We emphasize that all the  parameters $(\tau,\xi,\theta,\varphi)$ can be conveniently estimated using the FFTs because of the unique property of ZC sequences -- a time-delayed ZC appears like one with a frequency offset.


\subsubsection{Refined Channel Estimation Using Newton's Method}
Given the coarse parameter estimation, we use Newton's iterative method for refined estimation of $\zeta,\xi$ and $\theta,\varphi$.

With $\hat{\theta}$ and $\hat{\varphi}$ being fixed, since $\zeta$ and $\xi$ are independent of the denominator of (\ref{eq.MLnew1}), $\zeta$ and $\xi$ can be estimated by
\ben\label{eq.zetaXiNewton}
\{\hat{\zeta},\hat{\xi}\}=\arg\max_{\zeta,\xi}
\left| \abf(\hat{\theta},\hat{\varphi})^H{\cal W}(\xi)^H\tilde{\Ybf}\dbf(-\zeta) \right|^2.
\een

To estimate $\theta$ and $\varphi$ with $\hat{\zeta}$ and  $\hat{\xi}$ being fixed, (\ref{eq.MLnew1}) can be simplified as
\ben\label{eq.angleNewton}
\{\hat{\theta},\hat{\varphi}\}=\arg\max_{\theta,\varphi}
\frac{\left| \abf(\theta,\varphi)^H\rbf \right|^2}
{\sum_{k=1}^{K}\left\Arrowvert\abf(\theta,\varphi)^H\Wbf(\Phi_k)^H\right\Arrowvert^2},
\een
where
\ben\label{eq.r}
\rbf\triangleq {\cal W}(\hat{\xi})^H\tilde{\Ybf}\dbf(-\hat{\zeta}).\een

Let $\Lambda_1,\Lambda_2$ denote the objective functions of (\ref{eq.zetaXiNewton}) and (\ref{eq.angleNewton}) respectively, i.e.,
\ben\label{eq.zetaXiNewton2}
\Lambda_1(\psibf_1)\triangleq
\left| \abf(\hat{\theta},\hat{\varphi})^H{\cal W}(\xi)^H\tilde{\Ybf}\dbf(-\zeta) \right|^2,
\een
and
\ben\label{eq.angleNewton2}
\Lambda_2(\psibf_2)\triangleq
\frac{\left| \abf(\theta,\varphi)^H\rbf \right|^2}
{\sum_{k=1}^{K}\left\Arrowvert\abf(\theta,\varphi)^H\Wbf(\Phi_k)^H\right\Arrowvert^2},
\een
where $\psibf_1\triangleq [\zeta,\xi]^T$ and $\psibf_2\triangleq [\theta,\varphi]^T$.

The derivation of the Hessian matrices and Jacobian vectors
\ben\label{eq.H_g1}
\Hbf _1 = \frac{\partial^2\Lambda_1 }{\partial \psibf_1 \partial \psibf_1^T},
\;
\gbf_1 = \frac{\partial\Lambda_1}{\partial \psibf_1},
\een
and
\ben\label{eq.H_g2}
\Hbf _2 = \frac{\partial^2\Lambda_2 }{\partial \psibf_2 \partial \psibf_2^T},
\;
\gbf_2 = \frac{\partial\Lambda_2 }{\partial \psibf_2}
\een
are routine and are thus relegated to Appendix A in the supplementary document online https://github.com/csrlab-fudan/ris-chan-est/blob/main/appendices.pdf.
We can update the estimation as
\ben\label{eq.newton1}
\psibf_1^{(i+1)}=\psibf_1^{(i)}-s_1\Hbf_1^{-1}\gbf_1,
\een
and
\ben\label{eq.newton2}
\psibf_2^{(i+1)}=\psibf_2^{(i)}-s_2\Hbf_2^{-1}\gbf_2,
\een
where $s_1, s_2$ are the step sizes determined by the backtracking line search method \cite{boyd2004convex} to maximize (\ref{eq.zetaXiNewton2}) and (\ref{eq.angleNewton2}). 
Then a few alternations between (\ref{eq.zetaXiNewton2}) and (\ref{eq.angleNewton2}) are needed to achieve the final convergence.

In summary, the proposed scheme for the channel estimation based on RIS in the single-path scenario consists of two steps: i) to obtain the initial estimates according to (\ref{eq.MLnew2}); ii) to use Newton's method to search for the optimal estimation of (\ref{eq.zetaXiNewton2}) and (\ref{eq.angleNewton2}). Fast convergence is guaranteed owing to the good, albeit simple, initialization using the FFTs for the Newton's method.
At the end, the estimate of the complex-valued amplitude $\hat{\tilde{\beta}}$ is obtained by substituting the estimated parameters into (\ref{eq.betaTildeNew}).

We summarize the RIS channel estimation algorithm for the single-path scenario in Algorithm \ref{alg0}.
\begin{algorithm}[ht]
\caption{The RIS Channel Estimation in Single-path}\label{alg0}
\begin{algorithmic}[1]
\State{\bf Input:} the received signal $\Ybf$.
\State Obtain coarse solution $\{\hat{\zeta},\hat{\xi},\hat{\theta},\hat{\varphi}\}$ to (\ref{eq.MLnew2}) using FFTs as explained in Section \ref{subsec:fft};
\State Initialize $i=0$, $\psibf_1^{(0)}=\{\hat{\zeta},\hat{\xi}\}$ and $\psibf_2^{(0)}=\{\hat{\theta},\hat{\varphi}\}$;
\While {both $\psibf_1^{(i)}$ and $\psibf_2^{(i)}$ are not converged}
\State Calculate $\Hbf_1$, $\gbf_1$ according to (\ref{eq.H_g1});
\State Update $\psibf_1^{(i+1)}$ according to  (\ref{eq.newton1});
\State Update (\ref{eq.newton1});
\State Calculate $\Hbf_2$, $\gbf_2$ according to (\ref{eq.H_g2});
\State Update $\psibf_2^{(i+1)}$ according to (\ref{eq.newton2});
\State $i=i+1$;
\EndWhile
\State{\bf Output:} $\{\hat{\zeta},\hat{\xi},\hat{\theta},\hat{\varphi}\}$
\end{algorithmic}
\end{algorithm}

\subsection{Channel Estimation in The Multipath Case}\label{Hmulti}

Now we proceed to study the multipath case. The key idea is to use the classic SAGE method \cite{SAGE} to decompose the multipath problem to multiple single-path subproblems, to which the method in the previous section can be applied.

The SAGE consists of two steps : the expectation step (E-step), which calculates
\ben\label{eq.sageE}
\hat{\tilde{\Ybf}}_u
=\tilde{\Ybf}-\sum_{\substack{u^\prime=1\\ u^\prime\neq u}}^U\tilde{\Ybf}_{u^\prime},
\een
where $\tilde{\Ybf}$ is as defined in (\ref{eqYtilde}) and
\ben\label{eq.sageE2}
\tilde{\Ybf}_u = \hat{\tilde{\beta}}_u {\cal W}(\hat{\xi}_u)\abf(\hat{\theta}_u,\hat{\varphi}_u)\dbf(\hat{\zeta}_u) ^T;
\een
the maximization step (M-step), which calculates
\ben\begin{split} \label{eq.sageM}
&\{\hat{\zeta}_u,\hat{\xi}_u,\hat{\theta}_u,\hat{\varphi}_u\}\\
&=\arg\max_{\zeta_u,\xi_u,\theta_u,\varphi_u} \frac{\left| \abf(\theta_u,\varphi_u)^H{\cal W}(\xi_u)^H\hat{\tilde{\Ybf}}_u\dbf(-\zeta_u) \right|^2}{\sum_{k=1}^{K}\left\Arrowvert\abf(\theta_u,\varphi_u)^H\Wbf(\Phi_k)^H\right\Arrowvert^2},
\end{split}\een
to which the solution is delineated in the previous subsection, and
\ben\label{eq.sageM2}
\hat{\tilde{\beta}}_u=\frac{\abf(\hat{\theta}_u,\hat{\varphi}_u)^H{\cal W}(\hat{\xi}_u)^H\hat{\tilde{\Ybf}}_u\dbf(-\hat{\zeta}_u)}
{L\times{\sum_{k=1}^{K}\left\Arrowvert\abf(\hat{\theta}_u,\hat{\varphi}_u)^H\Wbf(\Phi_k)^H\right\Arrowvert^2}}.
\een

We refer to the consecutive iterations for updating the parameter estimates of all the multipaths for one round as an iteration cycle of the SAGE algorithm.

To initialize the SAGE procedure, we assume the received signal contains only one path and estimate $\{\hat{\zeta}_1,\hat{\xi}_1,\hat{\theta}_1,\hat{\varphi}_1,\hat{\tilde{\beta}}_1\}^{(1)}$ via (\ref{eq.sageM}) and (\ref{eq.sageM2}), where the superscript denotes the iteration index; thus $\tilde{\Ybf}_1^{(1)} = \hat{\tilde{\beta}}_1 {\cal W}(\hat{\xi}_1)\abf(\hat{\theta}_1,\hat{\varphi}_1)\dbf(\hat{\zeta}_1) ^T$ is obtained. Then we set $\hat{\tilde{\Ybf}}_2^{(1)}=\tilde{\Ybf} -\tilde{\Ybf} _1^{(1)}$ and estimate $\{\hat{\zeta}_2,\hat{\xi}_2,\hat{\theta}_2,\hat{\varphi}_2,\hat{\tilde{\beta}}_2 \}^{(1)}$ via (\ref{eq.sageM}) and (\ref{eq.sageM2}); thus $\tilde{\Ybf}_2^{(1)}$ is obtained by (\ref{eq.sageE2}).
Next we set $\hat{\tilde{\Ybf} }_3^{(1)}=\tilde{\Ybf} -\tilde{\Ybf}_1^{(1)}-\tilde{\Ybf}_2^{(1)}$ and estimate $\{\hat{\zeta}_3,\hat{\xi}_3,\hat{\theta}_3,\hat{\varphi}_3,\hat{\tilde{\beta}}_3 \}^{(1)}$,  and so on. This procedure continues until $\{\hat{\zeta}_U,\hat{\xi}_U,\hat{\theta}_U,\hat{\varphi}_U,\hat{\tilde{\beta}}_U\}^{(1)}$ with
$\hat{\tilde{\Ybf} }_U^{(1)}=\tilde{\Ybf}-\tilde{\Ybf} _1^{(1)}-\cdots-\tilde{\Ybf} _{U-1}^{(1)}$.

In the second round of iteration, first estimate $\{\hat{\zeta}_1,\hat{\xi}_1,\hat{\theta}_1,\hat{\varphi}_1,\hat{\tilde{\beta}}_1\}^{(2)}$ according to (\ref{eq.sageM}) and (\ref{eq.sageM2}) with $\hat{\tilde{\Ybf}}_1^{(2)}=\tilde{\Ybf} -\tilde{\Ybf} _2^{(1)}-\tilde{\Ybf }_3^{(1)}-\cdots-\tilde{\Ybf} _{U}^{(1)}$, and obtain $\tilde{\Ybf}_1^{(2)}$ by (\ref{eq.sageE2}). Then estimate $\{\hat{\zeta}_2,\hat{\xi}_2,\hat{\theta}_2,\hat{\varphi}_2,\hat{\tilde{\beta}}_2\}^{(2)}$ according to (\ref{eq.sageM}) and (\ref{eq.sageM2}) with $\hat{\tilde{\Ybf}}_2^{(2)}=\tilde{\Ybf} -\tilde{\Ybf} _1^{(2)}-\tilde{\Ybf }_3^{(1)}-\cdots-\tilde{\Ybf} _{U}^{(1)}$, and obtain $\tilde{\Ybf}_2^{(2)}$ by (\ref{eq.sageE2}) and so on. Proceed the iterations to update the parameters of each path until convergence. The whole SAGE iteration process is summarized in Algorithm \ref{alg1}.

In the above derivations we assume that the number of multipaths $U$ is known. In practice, it can be estimated using, e.g., the Akaike's  information criterion (AIC) \cite{AIC}. 

\begin{algorithm}[ht]
\caption{The RIS Channel Estimation in Multipath}\label{alg1}
\begin{algorithmic}[1]
\State{\bf Input:}  the number of multipath  $U$ and the received signal $\Ybf$.
\State Initialize $\tilde{\Ybf}=\Ybf\diag(\sbf^*), \tilde{\Ybf}_1=\tilde{\Ybf}_2=\cdots=\tilde{\Ybf}_U=0$;
\While {not converged}
\For {$u=1:U$}
\State Calculate $\hat{\tilde{\Ybf}}_u$ according to (\ref{eq.sageE});
\State Calculate $\{\hat{\zeta}_u,\hat{\xi}_u,\hat{\theta}_u,\hat{\varphi}_u\}$ according to (\ref{eq.sageM}) using Algorithm \ref{alg0};
\State Calculate $\hat{\tilde{\beta}}_u$ according to (\ref{eq.sageM2});
\State Calculate $\tilde{\Ybf}_u$ according to (\ref{eq.sageE2});
\EndFor
\EndWhile
\State{\bf Output:} $\{\hat{\zetabf},\hat{\xibf},\hat{\thetabf},\hat{\varphibf},\hat{\tilde{\betabf}} \}$
\end{algorithmic}
\end{algorithm}

According to (\ref{betaT}) and (\ref{eq.interPoint}), we can get
\ben
\hat{\beta}_u=\hat{\tilde{\beta}}_u/e^{\frac{j\pi \hat{\tau}_u^2}{\tilde{L}}},
\een
and
\ben
\hat{\tau}_u=(\hat{\xi}_u-\hat{\zeta}_u)\tilde{L}.
\een
After obtaining $\hat{\taubf}=[\hat{\tau}_1,\cdots,\hat{\tau}_U]^T$ and $\hat{\betabf}=[\hat{\beta}_1,\cdots,\hat{\beta}_U]^T$,
 we obtain the {\em estimated} time-domain channel response at time $0$
\ben\label{eq.H_ris}
\hat{\Hbf}^{(0)}=\sum_{u=1}^{U}\hat{\beta}_u\abf(\hat{\theta}_u,\hat{\varphi}_u) p(\tau-\hat{\tau}_u).
\een
Hence the {\em projected} time-domain channel after time $T$ is
\ben\label{eq.H_risproj}
\hat{\Hbf}^{(T)}=\sum_{u=1}^{U}\hat{\beta}_ue^{j2\pi\hat{\xi}_u T} \abf(\hat{\theta}_u,\hat{\varphi}_u) p(\tau-\hat{\tau}_u).
\een
where $p(\tau)$ is the raised cosine filter pulse shaper
\ben
p(\tau)={\rm sinc}(\tau)\frac{{\rm cos}(\pi\mu \tau)}{1-(2\mu \tau)^{2}},
\een
where we have assumed the Nyquist sampling interval $T_s=1$ for notational simplicity and $\mu$ is the roll-off factor.

Note that the channel projection is made possible by the estimation of the Doppler frequency offsets $\hat{\xi}_u, u=1,\ldots, U$.

The frequency-domain channel response of the estimated channel and the projected channel from UE to RIS can be obtained by applying a $\tilde{L}$-point FFT to (\ref{eq.H_ris}) or (\ref{eq.H_risproj}).

\subsection{Cramer Rao Bound}\label{subsectionCRB}
To gauge the effectiveness of the proposed algorithm, we derive the CRB of the channel parameter estimation as follows: 

Define all the unknown parameters as $\Omegabf=[\taubf,\xibf,\thetabf,\varphibf,\betabf_r,\betabf_i]\in\mathbb{R}^{6U}$, where $\taubf=[\tau_1,\cdots,\tau_U]^T$,  $\xibf=[\xi_1,\cdots,\xi_U]^T$, $\thetabf=[\theta_1,\cdots,\theta_U]^T$, $\varphibf=[\varphi_1,\cdots,\varphi_U]^T$, $\betabf_{r}=[\Re\{\beta_1\},\cdots,\Re\{\beta_U\}]^T$ and $\betabf_{i}=[\Im\{\beta_1\},\cdots,\\\Im\{\beta_U\}]^T$. Denote $\ybf\triangleq\vec(\Ybf)$, where 
\begin{align}
\Ybf = \sum_{u=1}^{U}{\cal W}(\xi_u)\abf(\theta_u,\varphi_u)\beta_u(\xbf(\tau_u)\odot \dbf(\xi_u))^T+\Zbf.
\end{align}

Since $\vec(\Abf\Bbf\Cbf^T)=(\Cbf\otimes\Abf)\vec(\Bbf)$, we have
\begin{align}
&\ybf=\bbf+\vec(\Zbf)
\nonumber\\
&=\sum_{u=1}^U\beta_u\left[\xbf(\tau_u)\odot \dbf(\xi_u)\right]\otimes\left[{\cal W}(\xi_u)\abf(\theta_u,\varphi_u)\right]+\vec(\Zbf).
\end{align}

Given that $\Zbf$ is white Gaussian noise, $\ybf\sim {\cal CN}(\bbf,\sigma^2\Ibf)$, the expression of the fisher information matrix (FIM) can be written as
\ben\label{eq.FimMatrix}
\Fbf
=\frac{2}{\sigma^2}\Re\left\{\frac{\partial\bbf^{H}}{\partial\Omegabf^T}
\frac{\partial\bbf}{\partial\Omegabf}\right\},
\een
where $\frac{\partial\bbf}{\partial\Omegabf}=\left[\frac{\partial\bbf}{\partial\taubf},\frac{\partial\bbf}{\partial\xibf},\frac{\partial\bbf}{\partial\thetabf},\frac{\partial\bbf}{\partial\varphibf},
\frac{\partial\bbf}{\partial\betabf_r},\frac{\partial\bbf}{\partial\betabf_i}\right]\in\mathbb{C}^{KM_rL\times6U}$.

For $\frac{\partial\bbf}{\partial\taubf}=\left[\frac{\partial\bbf}{\partial\tau_1},\cdots,\frac{\partial\bbf}{\partial\tau_U}\right]$, we have
\ben\label{eq.ytaud1}
\frac{\partial\bbf}{\partial\tau_u}
=\beta_u\left[\frac{\partial\xbf(\tau_u)}{\partial\tau_u}\odot \dbf(\xi_u)\right]
\otimes\left[{\cal W}(\xi_u)\abf(\theta_u,\varphi_u)\right].
\een

For $\frac{\partial\bbf}{\partial\xibf}=\left[\frac{\partial\bbf}{\partial\xi_1},\cdots,\frac{\partial\bbf}{\partial\xi_U}\right]$, we have
\ben\begin{split}\label{eq.yxid1}
&\frac{\partial\bbf}{\partial\xi_u}
=\beta_u\left[\xbf(\tau_u)\odot \frac{\partial\dbf(\xi_u)}{\partial\xi_u}\right]
\otimes\left[{\cal W}(\xi_u)\abf(\theta_u,\varphi_u)\right]\\
&+\beta_u\left[\xbf(\tau_u)\odot \dbf(\xi_u)\right]
\otimes\left[\frac{\partial{\cal W}(\xi_u)}{\partial\xi_u}\abf(\theta_u,\varphi_u)\right].
\end{split}\een

For $\frac{\partial\bbf}{\partial\thetabf}=\left[\frac{\partial\bbf}{\partial\theta_1},\cdots,\frac{\partial\bbf}{\partial\theta_U}\right]$, we have
\begin{align}\label{eq.ythetad1}
\frac{\partial\bbf}{\partial\theta_u}
=\beta_u\left[\xbf(\tau_u)\odot \dbf(\xi_u)\right]
\otimes\left[{\cal W}(\xi_u)\frac{\partial\abf(\theta_u,\varphi_u)}{\partial\theta_u}\right].
\end{align}

For $\frac{\partial\bbf}{\partial\varphibf}=\left[\frac{\partial\bbf}{\partial\varphi_1},\cdots,\frac{\partial\bbf}{\partial\varphi_U}\right]$, we have
\begin{align}\label{eq.yphid1}
\frac{\partial\bbf}{\partial\varphi_u}
=\beta_u\left[\xbf(\tau_u)\odot \dbf(\xi_u)\right]
\otimes\left[{\cal W}(\xi_u)\frac{\partial\abf(\theta_u,\varphi_u)}{\partial\varphi_u}\right].
\end{align}

For $\frac{\partial\bbf}{\partial\betabf_{r}}=\left[\frac{\partial\bbf}{\partial\Re\{\beta_1\}},\cdots,\frac{\partial\bbf}{\partial\Re\{\beta_U\}}\right]$, we have
\ben\label{eq.ybetad1}
\frac{\partial\bbf}{\partial\Re\{\beta_u\}}
=\left[\xbf(\tau_u)\odot \dbf(\xi_u)\right]\otimes\left[{\cal W}(\xi_u)\abf(\theta_u,\varphi_u)\right].
\een

Similarly, $\frac{\partial\bbf}{\partial\betabf_{i}}=\sqrt{-1}\frac{\partial\bbf}{\partial\betabf_{r}}$.

Based on the above derivations, we insert $\frac{\partial\bbf}{\partial\Omegabf}$ into (\ref{eq.FimMatrix}). The CRBs of $\left\{\tau_u,\xi_u,\theta_u,\varphi_u\right\}$ are given by
\ben\label{eq.crbtau}
{\rm CRB}(\tau_u)=[\Fbf^{-1}]_{u,u},
\een
\ben\label{eq.crbxi}
{\rm CRB}(\xi_u)=[\Fbf^{-1}]_{U+u,U+u},
\een
\ben
{\rm CRB}(\theta_u)=[\Fbf^{-1}]_{2U+u,2U+u},
\een
\ben
{\rm CRB}(\varphi_u)=[\Fbf^{-1}]_{3U+u,3U+u}.
\een
\ben
{\rm CRB}(\Re{(\beta_u)})=[\Fbf^{-1}]_{4U+u,4U+u},
\een
and
\ben
{\rm CRB}(\Im{(\beta_u)})=[\Fbf^{-1}]_{5U+u,5U+u}.
\een

\section{Estimation of the RIS Channel With the Direct Link} \label{sec4}

This section addresses the more general RIS channel estimation problem in the presence of direct path between the BS and the UE. Using the RIS property that the RIS can independently reflect the incident signal by controlling the reflection amplitude \cite{RIS_switching}, we propose to estimate the channel by two steps: i) to shut down the RIS, and estimate $\Hbf_d$ in a way similar to that given in Section $\textrm{III}$; ii) to turn on the RIS, and we can get the received signal of the reflected path after subtracting the received signal of the direct link, since $\Hbf_d$ has been estimated from the first step.

\subsection{The Channel Estimation of $\Hbf_d$}\label{subsection:Hd}
With the RIS being shut down, the user-to-BS channel is a $D$-tap ISI channel, in which the BS receives
\begin{align}
\ybf(n)=\sum_{d=1}^{D}\alpha_d \cbf(\overline{\theta}_d) x(n-\overline{\tau}_d)e^{j2\pi\overline{\xi}_d n}+\zbf(n).
\end{align}
Similar to (\ref{eq.MLOri4}), the channel parameters can be estimated as
\begin{align}\label{eq.MLOrid}
\{\hat{\tilde{\alphabf}},\hat{\overline{\zetabf}},\hat{\overline{\xibf}},\hat{\overline{\thetabf}}\}\!=\!\arg\!\min_{\tilde{\alphabf},\overline{\zetabf},\overline{\xibf},\overline{\thetabf}}
\!\left\Arrowvert \tilde{\Ybf} - \sum_{d=1}^{D}{\cal \overline B}(\overline{\xi}_d)\cbf(\overline{\theta}_d)\tilde{\alpha}_d\dbf(\overline{\zeta}_d) ^T \right\Arrowvert^2_F,
\end{align}
where
\ben
{\mathcal{\overline B}}(\overline{\xi}_d)\triangleq\begin{bmatrix}\Ibf_{M_r} \\ \Ibf_{M_r}e^{j2\pi\overline{\xi}_d N} \\ \vdots \\ \Ibf_{M_r}e^{j2\pi\overline{\xi}_d(\overline K-1)N} \end{bmatrix}
\in{\mathbb C}^{\overline K M_r \times M_r}.
\een
$\overline K$ is the number of observations made for estimating the user-to-BS channel (in contrast, $K$ is the number of observations made for estimating the user-to-RIS channel).
Problem (\ref{eq.MLOrid}) can be similarly solved using the algorithms given in Section $\textrm{III}$ as explained in the next.

In the single-path case, (\ref{eq.MLOrid}) degenerates to be
\ben\begin{split}\label{eq.ML_d}
\{\hat{\tilde{\alpha}},\hat{\overline{\zeta}},\hat{\overline{\xi}},\hat{\overline{\theta}}\}
=\arg\min_{\tilde{\alpha},\overline{\zeta},\overline{\xi},\overline{\theta}}\left\Arrowvert \tilde{\Ybf} - \tilde{\alpha} {\cal{\overline B}} (\overline{\xi})\cbf(\overline{\theta})\dbf(\overline{\zeta})^T\right\Arrowvert_F,
\end{split}\een
where$\tilde{\Ybf}=\Ybf\diag(\sbf^*)$, $\Ybf$ is the received signal of the direct path after $\overline K$ observations.

Expanding the cost function of (\ref{eq.ML_d}) to obtain the quadratic function of $\tilde{\alpha}$, since ${\pbf}(\overline{\xi})$, $\dbf(\overline{\zeta})$ and $\cbf(\overline{\theta})$ all have unit-modulus elements, we can minimize the cost function with respect to $\tilde{\alpha}$ to obtain
\begin{align}\label{eq.alphaTildeNew}
&\tilde{\alpha}=\frac{\cbf(\overline{\theta})^H{\cal \overline B}(\overline{\xi})^H\tilde{{\Ybf}}\dbf(-\overline{\zeta})}
{L\overline KM_r}.
\end{align}

Inserting (\ref{eq.alphaTildeNew}) into (\ref{eq.ML_d}) yields
\begin{align}\label{eq.ML_dnew}
&\{\hat{\overline{\zeta}},\hat{\overline{\xi}},\hat{\overline{\theta}}\}
=\arg\max_{\overline{\zeta},\overline{\xi},\overline{\theta}} {\left| \cbf(\overline{\theta})^H{\cal \overline B}(\overline{\xi})^H\tilde{{\Ybf}}\dbf(-\overline{\zeta}) \right|}
\nonumber\\
&=\arg\max_{\overline{\zeta},\overline{\xi},\overline{\theta}} {\left| \sum_{k=1}^{\overline K}\cbf(\overline{\theta})^H\tilde{{\Ybf}}_k\dbf(-\overline{\zeta})e^{-j2\pi N\overline{\xi} (k-1)} \right|},
\end{align}
which can be coarsely solved using FFTs as explained in the following.
\subsubsection{Coarse Solution to (\ref{eq.ML_dnew}) Using FFTs}
According to the expression $\cbf(\overline{\theta})$ given in (\ref{eq:cbf}), $\cbf(\overline{\theta})^H\tilde{\Ybf}_k$ can be evaluated efficiently over the mesh grids $-\frac{\bar{d}_x}{\lambda}\cos{\overline{\theta}} = 0, \frac{1}{N_{\overline{\theta}}}, \ldots, \frac{N_{\overline{\theta}}-1}{N_{\overline{\theta}}}$ using $N_{\overline{\theta}}$-point FFTs to obtain
 \ben \label{eqCbf}
\Cbf^{(k)}\triangleq\begin{bmatrix}
\begin{array}{c}\vdots\\ \cbf(\overline{\theta})^H\tilde{\Ybf}_k \\ \vdots \end{array}
\end{bmatrix}\in {\mathbb C}^{N_{\overline{\theta}} \times L},
\een
According to the expression $\dbf(\cdot)$ given in (\ref{eq.xtauxi}),
$\Cbf^{(k)}\dbf(-\overline{\zeta})$ can be evaluated efficiently over the mesh grids $\overline{\zeta} = 0, \frac{1}{N_{\overline{\zeta}}}, \ldots, \frac{N_{\overline{\zeta}}-1}{N_{\overline{\zeta}}}$ using $N_{\overline{\zeta}}$-point FFTs to obtain
 \ben \label{eqVbf}
\Vbf^{(k)}\triangleq\begin{bmatrix}\Cbf^{(k)}\dbf(0), \cdots, \Cbf^{(k)}\dbf(-\frac{N_{\overline{\zeta}}-1}{N_{\overline{\zeta}}}) \end{bmatrix}\in {\mathbb C}^{N_{\overline{\theta}}\times N_{\overline{\zeta}}}.
\een
Thus, we can evaluate (\ref{eq.ML_dnew}) by applying $N_{\overline{\xi}}$-point FFTs across the $\overline K$ matrices $\Vbf^{(k)}$ to obtain a 3D tensor $\Pbf \in {\mathbb C}^{N_{\overline{\theta}}\times  N_{\overline{\zeta}} \times N_{\overline{\xi}}}$, i.e.,
 \ben \label{eqPfft}
 \Pbf(:,:,n_{\overline{\xi}})=
 \sum_{k=1}^{\overline K}\Vbf^{(k)}
 e^{-j2\pi \frac{(n_{\overline{\xi}}-1)(k-1)}{N_{\overline{\xi}} }}, n_{\overline{\xi}}=1,2,\ldots, N_{\overline{\xi}}.
 \een

Thus, (\ref{eq.ML_dnew}) can be evaluated by computing the absolute values of every entries of $\Pbf$, which we refer to as $|\Pbf|$.
By identifying the indices of the largest entry of $|\Pbf|$, we obtain the estimation $\hat{\overline{\zeta} },\hat{\overline{\xi} },\hat{\overline{\theta}}$ using the procedure similar to that of (\ref{eqffff})-(\ref{FFTpara}).
\subsubsection{Refined Channel Estimation}
Given the initial estimation, we use Newton's iterative method for refined estimation of $\overline{\zeta}$, $\overline{\xi}$, and $\overline{\theta}$.

Let $\Lambda_3$ denote the squared objective function of (\ref{eq.ML_dnew}), i.e.,
\ben\label{eq.dNewton}
\Lambda_3(\psibf_3)\triangleq
\left\arrowvert \cbf(\overline{\theta})^H{\cal \overline B}(\overline{\xi})^H\tilde{{\Ybf}}\dbf(-\overline{\zeta}) \right\arrowvert^2,
\een
where $\psibf_3\triangleq [\overline{\zeta}, \overline{\xi}, \overline{\theta}]^T$.
The routine derivations of the Hessian matrix and Jacobian vector
\ben\label{eq.H_g3}
\Hbf _3 = \frac{\partial^2\Lambda_3 }{\partial \psibf_3 \partial \psibf_3^T},
\;
\gbf_3 = \frac{\partial\Lambda_3}{\partial \psibf_3}
\een
can be found in Appendix B in the supplementary document online https://github.com/csrlab-fudan/ris-chan-est/blob/main/appendices.pdf.
we can update the estimation as
\ben\label{eq.newton3}
\psibf_3^{(i+1)}=\psibf_3^{(i)}-s_3\Hbf_3^{-1}\gbf_3,
\een
where $s_3$ are the step sizes determined by the backtracking line search method to maximize (\ref{eq.dNewton}). Then the optimal channel estimation can be obtained by using Newton's method.

In the multipath case, the key idea to use the SAGE algorithm is similar to subsection \ref{Hmulti}. That is, to use the SAGE algorithm to decompose the multipath problem to multiple single-path subproblems.

The SAGE consists of two steps: the expectation step (E-step), which calculates
\ben\label{eq.sageEd1}
\hat{\tilde{{\Ybf}}}_d
=\tilde{{\Ybf}}-\sum_{\substack{d^\prime=1\\ d^\prime\neq d}}^D\tilde{{\Ybf}}_{d^\prime},
\een
where $\tilde{{\Ybf}}_d$ is defined as
\ben\label{eq.sageEd3}
\tilde{{\Ybf}}_d = \hat{\tilde{\alpha}}_d {\cal \overline B}(\hat{\overline{\xi}}_d)\cbf(\hat{\overline{\theta}}_d)\dbf(\hat{\overline{\zeta}}_d) ^T,
\een
and the  maximization step (M-step), which calculates
\ben\label{eq.sageMd1}
\{ \hat{\overline{\zeta}}_d,\hat{\overline{\xi}}_d,\hat{\overline{\theta}}_d \}\!=\!\arg\!\max_{ \overline{\zeta}_d,\overline{\xi}_d,\overline{\theta}_d }
\left| \cbf({\overline{\theta}}_d)^H{\cal \overline B}({\overline{\xi}}_d)^H\hat{\tilde{{\Ybf}}}_d\dbf(-{\overline{\zeta}}_d) \right|
\een
and
\ben\label{eq.sageMd2}
\hat{\tilde{\alpha}}_d=\frac{\cbf(\hat{\overline{\theta}}_d)^H{\cal \overline B}(\hat{\overline{\xi}}_d)^H\hat{\tilde{{\Ybf}}}_d\dbf(-\hat{\overline{\zeta}}_d)}{LKM_r}
\een
through $\hat{\tilde{\Ybf}}_d$.
The solution to (\ref{eq.sageMd1}) is already delineated above.
The whole SAGE procedure is summarized in Algorithm \ref{alg2}.
\begin{algorithm}[ht]
\caption{Channel Estimation  of $\Hbf_d$  in Multipath}\label{alg2}
\begin{algorithmic}[1]
\State{\bf Input:}  the number of multipath $D$ and the received signal ${\Ybf}$.
\State Initialize $\tilde{\Ybf}={\Ybf}\diag(\sbf^*), \tilde{\Ybf}_1=\tilde{\Ybf}_2=\cdots=\tilde{\Ybf}_D=0$;
\While {not converged}
\For {$d=1:D$}
\State Calculate $\hat{\tilde{\Ybf}}_d$ according to (\ref{eq.sageEd1});
\State Calculate $\{\hat{\overline{\zeta}}_d,\hat{\overline{\xi}}_d,\hat{\overline{\theta}}_d\}$ according to (\ref{eq.sageMd1});
\State Calculate $\hat{\tilde{\alpha}}_d$ according to  (\ref{eq.sageMd2})
\State Calculate $\tilde{\Ybf}_d$ according to (\ref{eq.sageEd3});
\EndFor
\EndWhile
\State{\bf Output:} $\{\hat{\overline{\zetabf}},\hat{\overline{\xibf}},\hat{\overline{\thetabf}},\hat{\tilde{\alphabf}} \}$
\end{algorithmic}
\end{algorithm}

According to (\ref{betaT}) and (\ref{eq.interPoint}), we can get
\ben
\hat{\alpha}_d=\hat{\tilde{\alpha}}_d/e^{\frac{j\pi\hat{\overline{\tau}}_d^2}{\tilde{L}}}, \;\;
\hat{\overline{\tau}}_d=(\hat{\overline{\xi}}_d-\hat{\overline{\zeta}}_d)\tilde{L}, \; d=1,\ldots, D.
\een
Then the {\em estimated} time-domain channel response from the UE to the BS is at time $0$ is
\ben\label{HdEst}
\hat{\Hbf}_{d}^{(0)}=\sum_{d=1}^{D}\hat{\alpha}_d e^{j2\pi\hat{\overline{\xi}}_d \overline{K}N} \cbf(\hat{\overline{\theta}}_d) p(\tau-\hat{\overline{\tau}}_d),
\een
and the {\em projected} time-domain channel response from the UE to the BS at time $T$ is
\ben\label{HdPro}
\hat{\Hbf}_{d}^{(T)}=\sum_{d=1}^{D}\hat{\alpha}_d e^{j2\pi\hat{\overline{\xi}}_d (T+\overline{K}N)} \cbf(\hat{\overline{\theta}}_d) p(\tau-\hat{\overline{\tau}}_d).
\een
The corresponding frequency-domain channel response can be obtained by applying an $\tilde{L}$-point FFT to (\ref{HdEst}) or (\ref{HdPro}).
\subsection{Estimation of the UE-to-RIS Channel}

Given the estimation of the parameters of the UE-to-BS direct-link channel $\hat{\overline{\zetabf}},\hat{\overline{\xibf}},\hat{\overline{\thetabf}},\hat{\tilde{\alphabf}}$, the channel information of the direct link after $\bar{K}$ observations can be obtained as follows
\ben\label{Ydot}
\tilde{\breve{\Ybf}}=\sum_{d=1}^{D} \hat{\tilde{\alpha}}_d e^{j2\pi\hat{\overline{\xi}}_d \overline KN} {\cal B}(\hat{\overline{\xi}}_d)\cbf(\hat{\overline{\theta}}_d)\dbf(\hat{\overline{\zeta}}_d) ^T,
\een
where ${\cal B}(\cdot)$ is as defined in (\ref{BMathcal}).

The SAGE consists of two steps : the expectation step (E-step), which calculates
\ben\label{eq.sageEt2}
\hat{\tilde{\Ybf}}_u
=\tilde{\Ybf}-\tilde{\breve{\Ybf}}-\sum_{\substack{u^\prime=1\\ u^\prime\neq u}}^U\tilde{\Ybf}_{u^\prime},
\een
and $\tilde{\Ybf}_u$ is defined in (\ref{eq.sageE2});
the maximization step (M-step), which calculates (\ref{eq.sageM}) and (\ref{eq.sageM2}) through $\hat{\tilde{\Ybf}}_u$.
The whole SAGE iteration process is summarized in Algorithm \ref{alg3}.
\begin{algorithm}[ht]
\caption{Estimation of the RIS Channel With the Direct Link $\Hbf_d$  in Multipath}\label{alg3}
\begin{algorithmic}[1]
\State{\bf Input:}  the number of multipath $D$, $U$ and the received signal.
\State Estimate $\{\hat{\overline{\zetabf}},\hat{\overline{\xibf}},\hat{\overline{\thetabf}},\hat{\tilde{\alphabf}} \}$ using Algorithm \ref{alg2};
\State Restruct $\tilde{\breve{\Ybf}}$ via (\ref{Ydot});
\State Initialize $\tilde{\Ybf}=\Ybf\diag(\sbf^*), \tilde{\Ybf}_1=\tilde{\Ybf}_2=\cdots=\tilde{\Ybf}_U=0$;
\While {not converged}
\For {$u=1:U$}
\State Calculate $\hat{\tilde{\Ybf}}_u$ according to (\ref{eq.sageEt2});
\State Calculate $\{\hat{\zeta}_u,\hat{\xi}_u,\hat{\theta}_u,\hat{\varphi}_u\}$ according to (\ref{eq.sageM}) using Algorithm \ref{alg0};
\State Calculate $\hat{\tilde{\beta}}_u$ according to (\ref{eq.sageM2});
\State Calculate $\tilde{\Ybf}_u$ according to (\ref{eq.sageE2});
\EndFor
\EndWhile
\State{\bf Output:} $\{\hat{\overline{\zetabf}},\hat{\overline{\xibf}},\hat{\overline{\thetabf}},\hat{\tilde{\alphabf}},\hat{\zetabf},\hat{\xibf},\hat{\thetabf},\hat{\varphibf},\hat{\tilde{\betabf}}  \}$
\end{algorithmic}
\end{algorithm}

Therefore, after the RIS is switched on, at time $n=0$, for example, the {\em estimated} and the {\em projected} time-domain channel responses from UE to RIS is as defined in (\ref{eq.H_ris}) and (\ref{eq.H_risproj}).
And the frequency-domain channel response can be obtained by applying FFTs to the time domain channels.

\section{Simulation Results}

This section evaluates the performance of the proposed scheme through numerical simulations.\footnote{The Matlab\textsuperscript{TM} codes used for generating the simulation results can be found online:\\ {https://github.com/csrlab-fudan/ris-chan-est/tree/main/chan-est-code}.}
Consider a $32\times 32$-element RIS (i.e., $M=PQ=1024$) with half wavelength inter-element spacing, and a BS with $M_r=6$ antennas.
The transmitted pilot is the ZC sequences of length $\tilde L=1024$ with cyclic prefix (CP) of length $T_{cp}= 64$.
The pulse shaper is a raised cosine filter with the roll-off factor $0.3$. The adjustable phases of the RIS are $\{0,\pi/2, \pi, 3\pi/2\}$, and the reflection efficiency of the RIS is 0.5.
The carrier frequency $f_c=28$GHz and the bandwidth $B=50$MHz. Hence the time-duration of an OFDM symbol is about 20$\mu s$. Out of the $1024$-length training sequence, $L=600$ samples corresponding to the lower frequency will be taken out for channel estimation.
The dimensions of the FFTs for solving (\ref{eq.MLnew2}) and (\ref{eq.ML_dnew}) are $ N_{\theta}=32$, $N_{\varphi}=32$, $N_{\overline{\theta}}=32$, $N_{\zeta}=N_{\overline{\zeta}}=1024$ and $N_{\xi}=N_{\overline{\xi}}=64$.
All the following simulation results are based on the average of 100 Monte-Carlo trials.


We  first simulate the single-path LOS scenario where the direct UE-to-BS link is blocked. The 2D-DOA $\theta=90^\circ,\varphi=60^\circ$, the Doppler frequency offset $\xi=3\times 10^{-6} /T_s$ (or 150Hz).  According to the velocity-Doppler
frequency offset translation $v=\frac{C\xi}{f_c}$, where $C$ is the speed of light, the corresponding moving speed $v\approx1.6m/s$. The time delay $\tau=0.5T_s$, and the channel complex gain $\beta=e^{j\phi}$, where $\phi$ was generated at random. The ZC training sequences are transmitted four times ($K=4$); for each of them the RIS phases are varied at random.
Fig. \ref{Fig:singlePathPara} shows the root mean square error (RMSE) estimation of the time delay, the Doppler frequency offset, and the two-dimensional angles versus the SNR. We can see that the RMSE results of the proposed scheme overlap with the CRBs as derived in Section \ref{subsectionCRB}, which verify the effectiveness of the proposed method.
\begin{figure}[ht]
\centering
{\psfig{figure={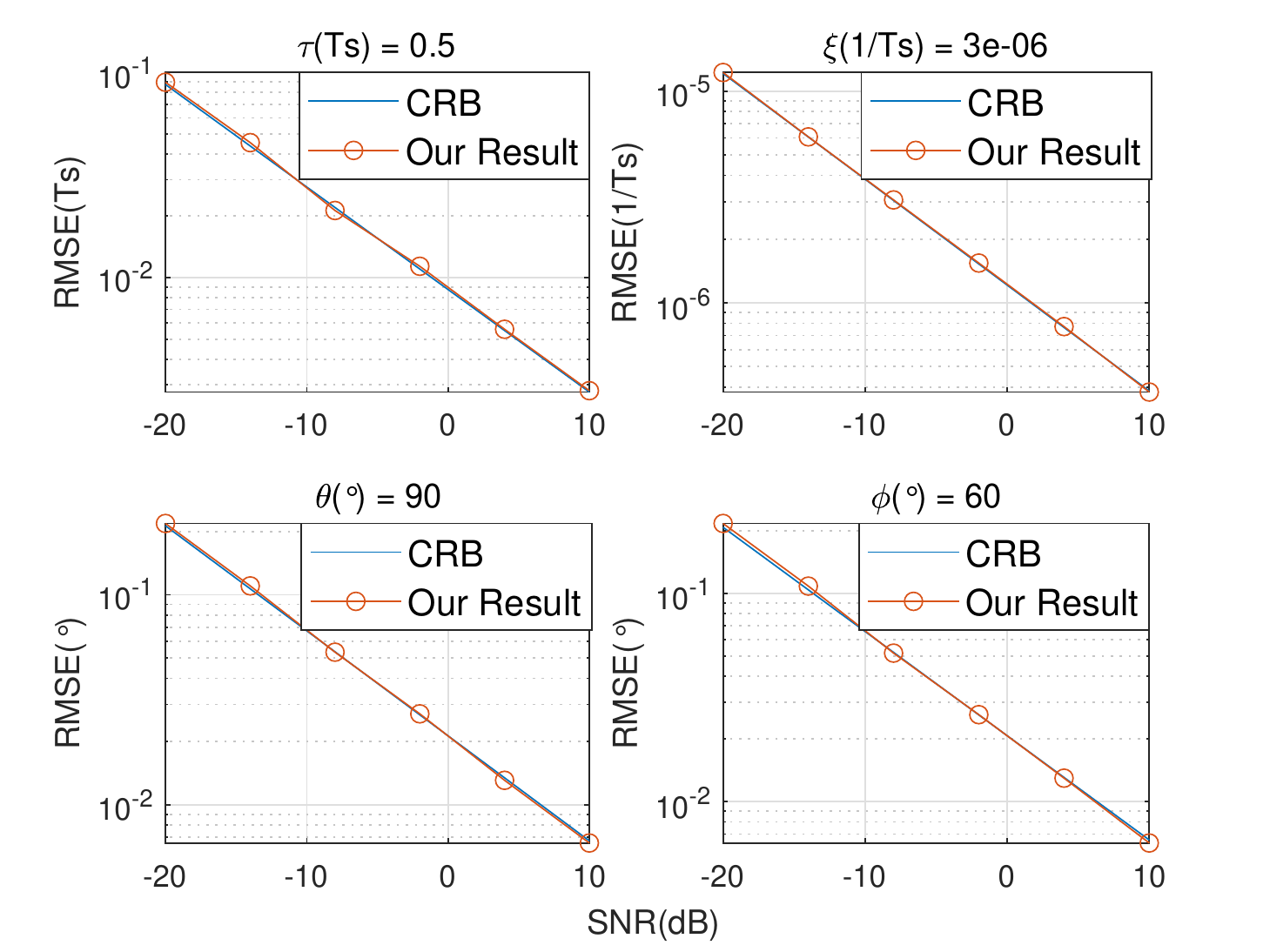} ,width=3.5in}}
\caption{The RMSEs of channel estimation and projection of a single-path case.
}
\label{Fig:singlePathPara}
\end{figure}


In the second example, the simulated channel has six paths with settings: $\taubf=[0.5,1.1,1.8,2.6,5.5,6.4]T_s$, $\xibf=[3\times10^{-4},-10^{-5},10^{-6},10^{-4},3\times10^{-6},-2\times10^{-5}]/T_s$ (the mobility corresponding to $\xi = 3\times10^{-4}$ is $v\approx 160.7$m/s), $\thetabf=[40^\circ,65^\circ,150^\circ,100^\circ,120^\circ,90^\circ]$, $\varphibf=[30^\circ,50^\circ,60^\circ,45^\circ,40^\circ,55^\circ]$ and the channel gains $[0.85e^{j\phi_1},0.8e^{j\phi_2}, 0.7e^{j\phi_3},0.65e^{j\phi_4},0.6e^{j\phi_5},0.5e^{j\phi_6}]$, where $\phi_1,\ldots,\phi_6$ are random. The ZC training sequences are transmitted six times ($K=6$), each with a different set of RIS phases. Fig. \ref{Fig:sixPath} shows the RMSE of the channel estimation and projection. The RMSE of the channel estimation/projection is calculated by
\ben\label{RMSE_RIS}
{\rm RMSE} = \sqrt{\frac{1}{\tilde{L}M} \|\hat{\Hbf}-\Hbf \|_F^2},
\een
where $\hat{\Hbf}$ and $\Hbf$ are the estimated/projected channel and the true channel from the UE to the RIS in the frequency domain, respectively, both with dimensionality $M\times \tilde{L}$. It can be seen from Fig. \ref{Fig:sixPath} that the channel projection is quite accurate after 10, 20 and even 40 OFDM symbols.
\begin{figure}[ht]
\centering
{\psfig{figure={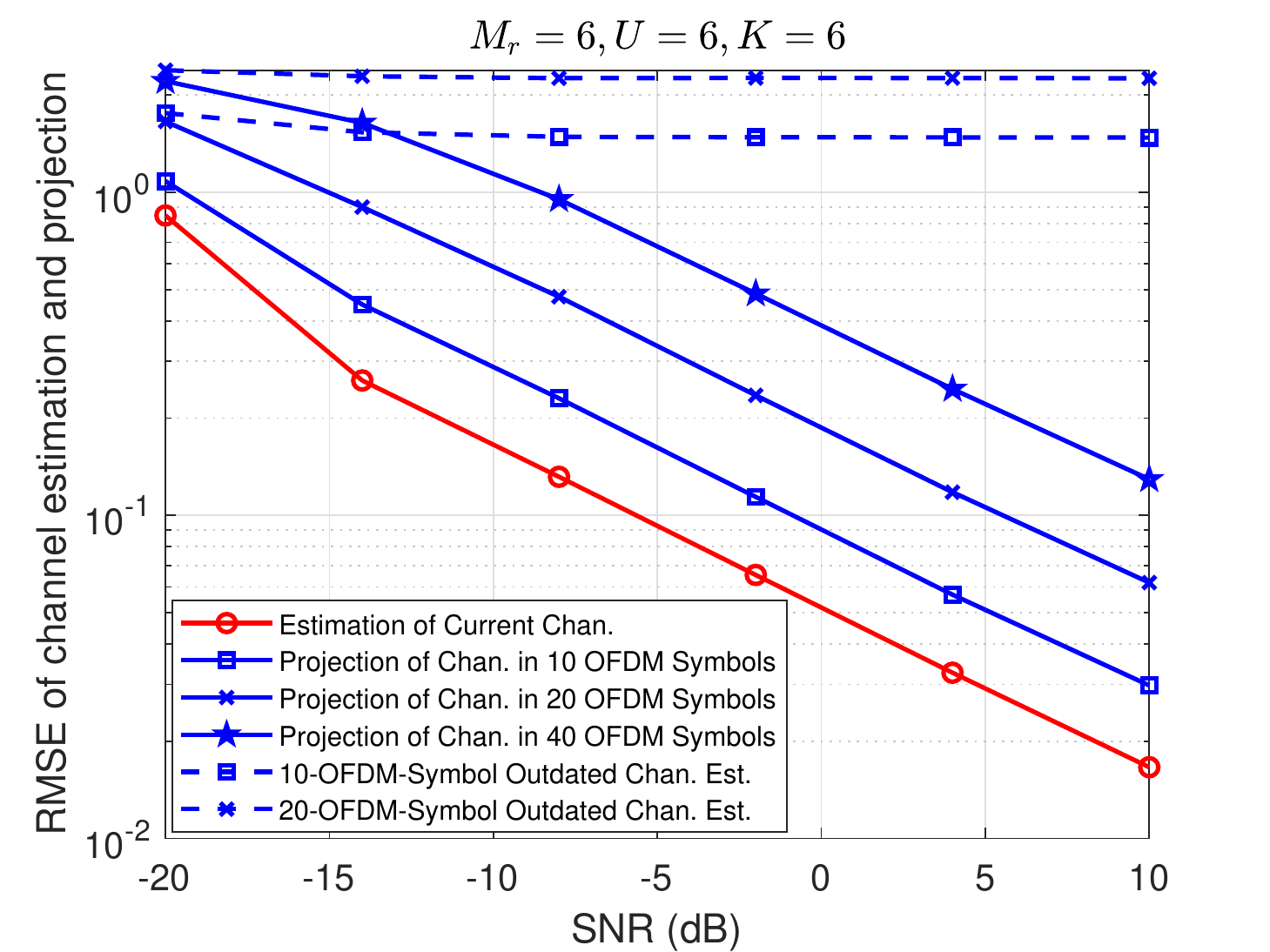} ,width=3.in}}
\caption{The RMSEs of channel estimation and projection of the UE-to-RIS channel with six paths.
}
\label{Fig:sixPath}
\end{figure}

The third example simulates the proposed scheme in the channel with the direct UE-to-BS link, where the  channel settings from UE to RIS are the same as those in the second example,  $\overline{\taubf}=[0.1,3.5]T_s$, $\overline{\xibf}=[6\times10^{-5},-9\times10^{-5}]/T_s$, $\overline{\thetabf}=[80^\circ,110^\circ]$, and the
channel gains $\alphabf= [e^{j\phi_7},0.8e^{j\phi_8}]$, where $\phi_7,\phi_8$ are randomly generated. To estimate the direct-link channel, the RIS is turned off at first and the UE transmits the ZC sequences four times ($\bar{K}=4$). Then the RIS is turned on with some random phases and the UE trains the ZC sequences six times ($K=6$). Fig. \ref{Fig:U6D2} (a) and Fig. \ref{Fig:U6D2} (b) show the RMSEs of the channel estimation and projection of $\Hbf$ and $\Hbf_d$ under different SNRs respectively.
The RMSE of the channel estimation/projection in Fig. \ref{Fig:U6D2} (a) is calculated by (\ref{RMSE_RIS}); the RMSE of the channel estimation/projection of the UE-to-BS direct link as shown in Fig. \ref{Fig:U6D2} (b) is similarly calculated by
\ben\label{RMSE_BS}
{\rm RMSE} = \sqrt{\frac{1}{\tilde{L}M_r} \|\hat{\Hbf}_d-\Hbf_d \|_F^2}.
\een

The circled lines in Fig. \ref{Fig:U6D2} show the RMSE performance of the channel estimation. The dashed lines represent the RMSE performance of the channel projection without Doppler offset, which show that the channel estimation becomes obsolete quickly after 10 or 20 OFDM symbols. The solid lines marked with squares, crosses and pentagram show that the channel projection enabled by the Doppler estimation can track the time-varying channel quite well after 10, 20 and even 40 OFDM symbols.
\begin{figure}[htb]
\centering
\subfloat[The RMSEs of channel estimation and projection of the UE-to-RIS channel $\Hbf$]{\includegraphics[width=3.in]{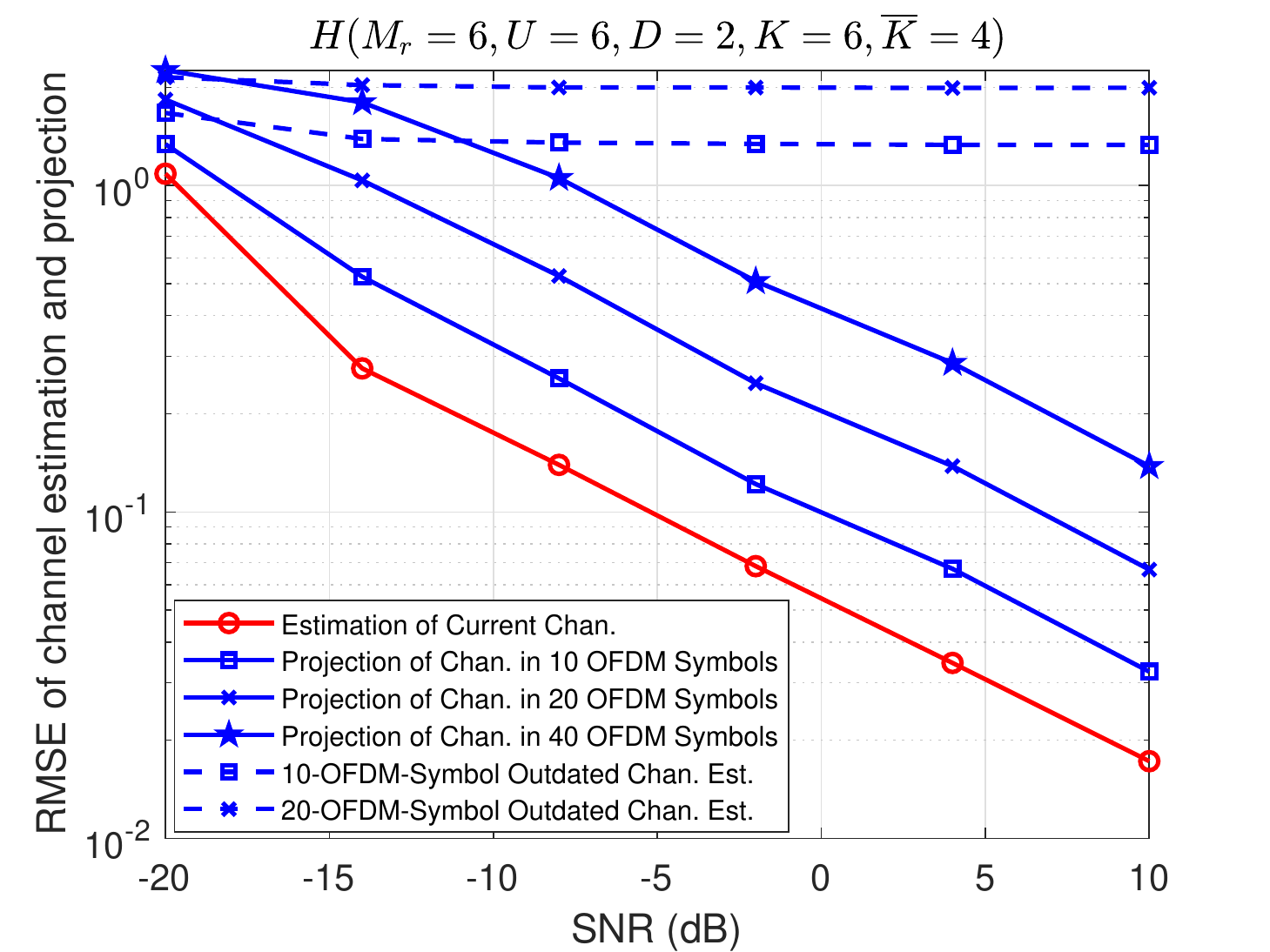}} \\
\subfloat[The RMSEs of channel estimation and projection of the UE-to-BS direct-link channel $\Hbf_d$]{\includegraphics[width=3.in]{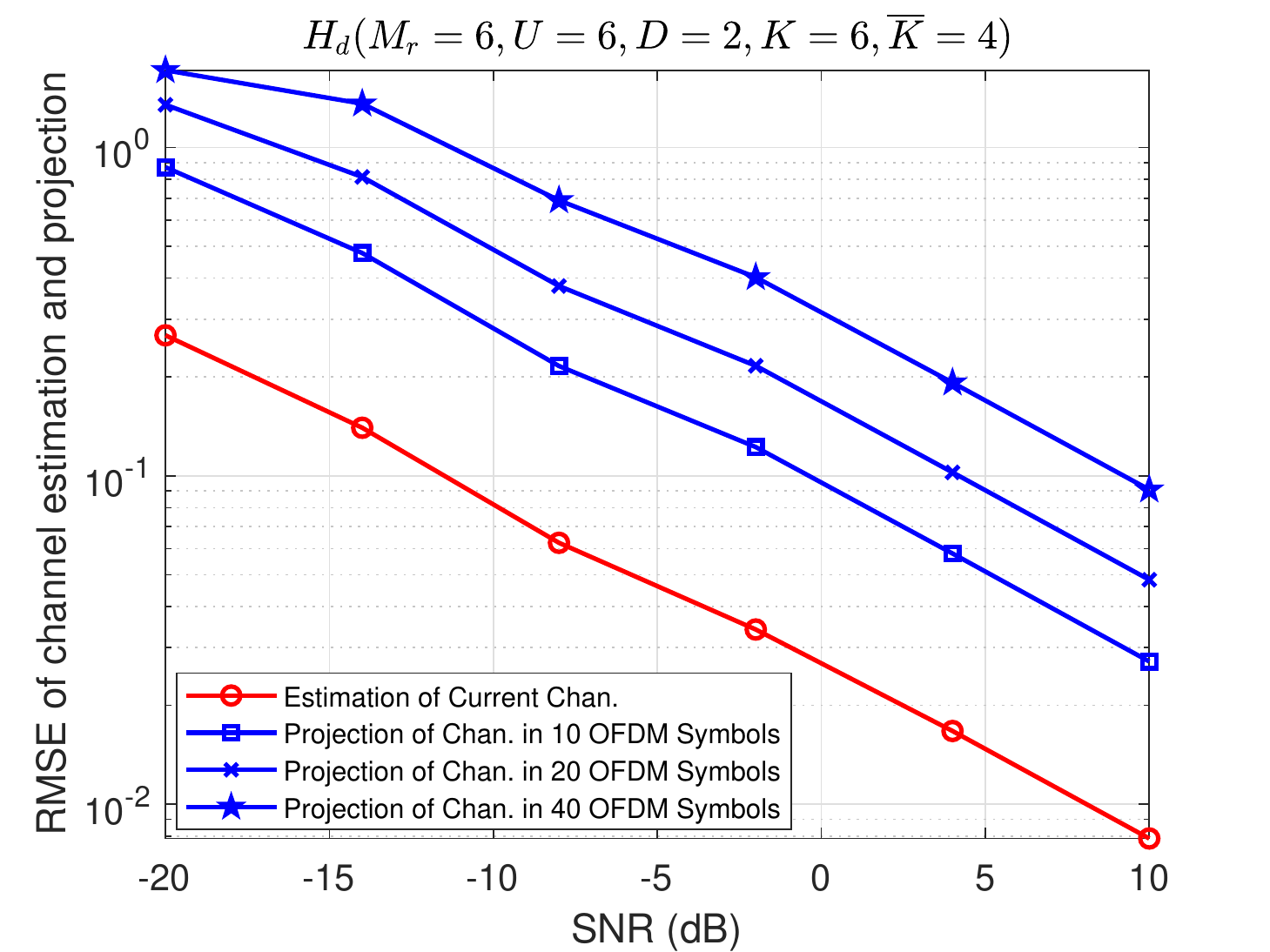}}
\caption{The RMSEs of channel estimation and projection of a RIS channel with direct link between the UE and the BS.}
\label{Fig:U6D2}
\end{figure}
The last example shows the RMSEs of channel estimation for different number of observations. The channel parameters are set as follows: $\taubf=[0.5,1.1,1.8,2.6]T_s$, $\xibf=[3\times10^{-4},-10^{-5},10^{-6},10^{-4}]/T_s$, $\thetabf=[40^\circ,65^\circ,150^\circ,100^\circ]$, $\varphibf=[30^\circ,50^\circ,60^\circ,45^\circ]$, and the channel gains $\betabf = [0.85e^{j\phi_1},0.8e^{j\phi_2},0.7e^{j\phi_3},0.6e^{j\phi_4}]$, where $\phi_1,\ldots,\phi_4$ are random.
The RMSEs of channel estimation of the $U=4$ case versus the number of observations $K=2,3,4,5,6,8,10$ are plotted in Fig. \ref{Fig:U4K}, which shows that it requires at least $K=3$ observations to estimate the channel parameters sufficiently well. It is not surprising that the accuracy of channel estimation improves as the number of observations increases. 
\begin{figure}[ht]
\centering
{\psfig{figure={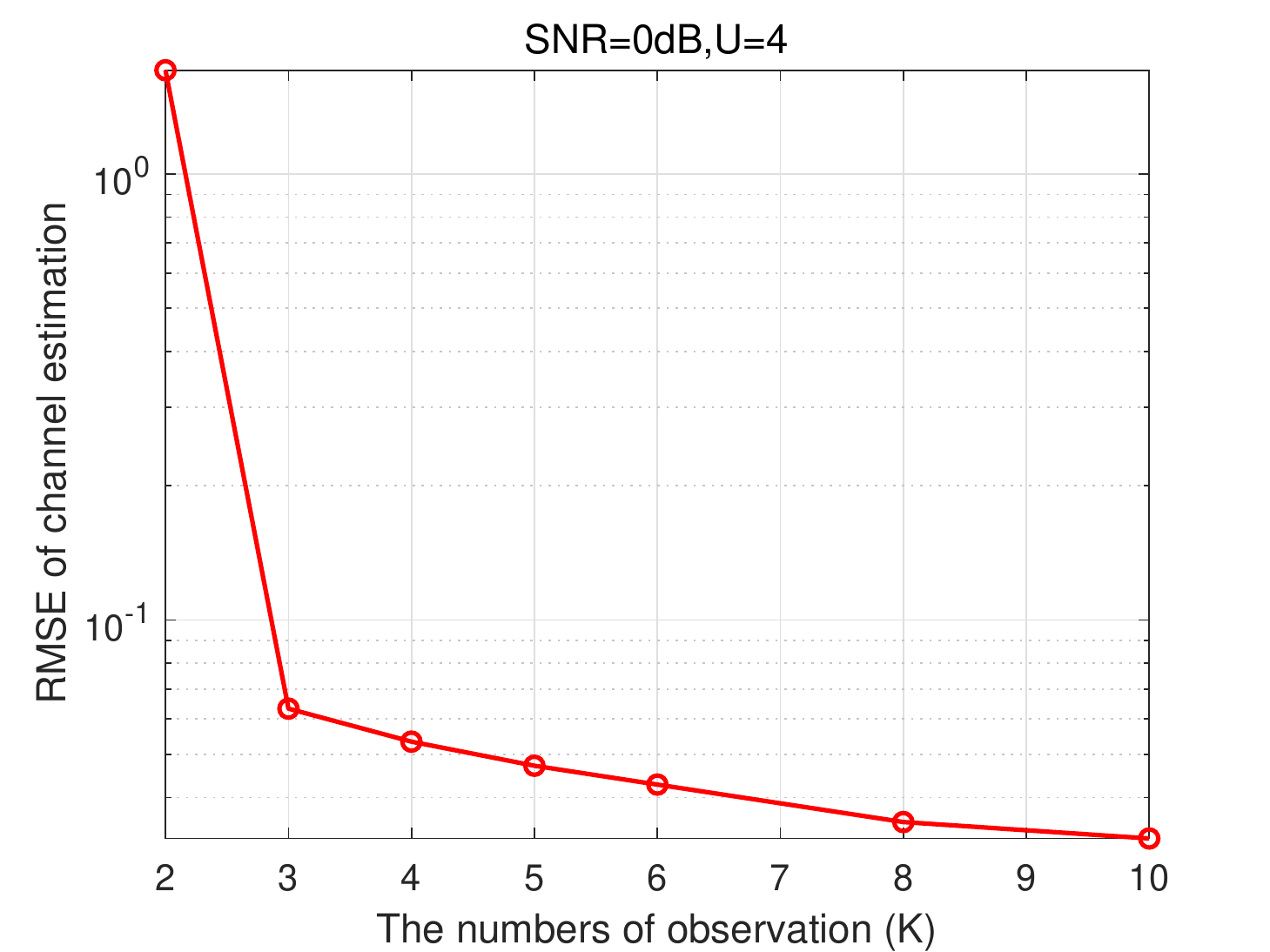} ,width=3.in}}
\caption{The RMSEs of channel estimation of a $U=4$ case in different $K$.}
\label{Fig:U4K}
\end{figure}

\section{Conclusion}
This paper studied uplink channel estimation for an RIS-assisted massive MIMO-OFDM system in a frequency selective channel. By parameterizing the channel in the directions of arrivals (DOA), the time delays, and the frequency offsets, we convert the channel estimation into a parameter estimation problem. We proposed to use the ZC pilot sequences and exploit the time delay-frequency offset ambiguity of the ZC sequences to simplify the high-dimensional parameter estimation problem. The multi-dimensional parameters can be first coarsely initialized via FFTs, before being fine-estimated using Newton's method. Base on the SAGE method, the multipath parameters estimation problem is transformed into multiple single-path problems. Owing to the estimation of the Doppler frequency offsets, the proposed algorithm can project the time-varying channel, which can greatly reduce the pilot overhead needed for channel estimation, and help mitigate the pilot contamination issue in massive MIMO communications.

\bibliographystyle{ieeetr}
\bibliography{ref}
\end{document}